\documentclass[preprint,tighten,showpacs,showkeys,nofootinbib,aps,amssymb,amsmath,floatfix]{revtex4}
\usepackage{psfig}
\usepackage{epsfig}
\usepackage{colordvi}
\usepackage{graphicx}

\def\lsim{\mathrel{\rlap{\lower4pt\hbox{\hskip1pt$\sim$}}
    \raise1pt\hbox{$<$}}}                
\def\gsim{\mathrel{\rlap{\lower4pt\hbox{\hskip1pt$\sim$}}
    \raise1pt\hbox{$>$}}}                

\def\kc{{{\buildrel \circ\over k}}}

\def\smn{{\sigma_{\mu\nu}}}
\newcommand{\be}{\begin{equation}}
\newcommand{\ee}{\end{equation}}
\newcommand{\bea}{\begin{eqnarray}} 
\newcommand{\eea}{\end{eqnarray}}

 \newcommand{\pslash}{{\not{\hspace{-0.08cm}p}}}  
 \newcommand{\ve}{\varepsilon}  
 \newcommand{\csw}{\, c_{\rm SW}}

\begin{document}
\setlength{\textheight}{23cm}
\textheight=23cm
\title{${\cal O}(a^2)$ corrections to the one-loop propagator and
  bilinears of clover fermions with Symanzik improved gluons}
\author{M. Constantinou$^a$, V. Lubicz$^b$, 
H. Panagopoulos$^a$, F. Stylianou$^a$}
\affiliation{
$^a$ Department of Physics, University of Cyprus,
P.O.Box 20537, Nicosia CY-1678, Cyprus\\
$^b$Dipartimento di Fisica, Universit\`a di Roma Tre and INFN, Sezione
di Roma Tre,
Via della Vasca Navale 84, I-00146 Roma, Italy\\
{\it email: }{\tt marthac@ucy.ac.cy, lubicz@fis.uniroma3.it,
  haris@ucy.ac.cy, fstyli01@ucy.ac.cy}}
\begin{abstract}
We calculate corrections to the fermion propagator and to
the Green's functions of all fermion bilinear operators of the form
$\bar\Psi \Gamma \Psi$, to one-loop in perturbation theory.

We employ the Wilson/clover action for fermions and the Symanzik
improved action for gluons.

The novel aspect of our calculations is that they are carried out to
second order in the lattice spacing, ${\cal O}(a^2)$. Consequently, they have 
addressed a number of new issues, most notably the appearance of loop
integrands with strong IR divergences (convergent only beyond 6
dimensions). Such integrands are not present in ${\cal O}(a^1)$ improvement
calculations; there, IR divergent terms are seen to have the same
structure as in the ${\cal O}(a^0)$ case, by virtue of parity under
integration, and they can thus be handled by well-known techniques. We
explain how to correctly extract the full ${\cal O}(a^2)$ dependence; in fact, our
method is generalizable to any order in $a$.

The ${\cal O}(a^2)$ corrections to the quark propagator and Green's
functions computed in this paper are useful to improve the
nonperturbative RI-MOM determination of renormalization constants for
quark bilinear operators. 

Our results depend on a large number of parameters: coupling constant,
number of colors, lattice spacing, external momentum, clover
parameter, Symanzik coefficients, gauge parameter. To make these
results most easily accessible to the reader, we have included them in
the distribution package of this paper, as an ASCII file named:
Oa2results.m ; the file is best perused as Mathematica input.
\end{abstract}

\medskip
\keywords{Lattice QCD, Lattice perturbation theory, Fermion
  propagator, Fermion bilinears, Improved actions} 
\medskip
\pacs{11.15.Ha, 12.38.Gc, 11.10.Gh, 12.38.Bx}

\maketitle

\section{Introduction}

A major issue facing Lattice Gauge Theory, since its early days, has
been the reduction of effects which are due to the finite lattice
spacing $a$, in order to better approach the elusive continuum
limit. A systematic framework to address this issue is Symanzik's
program \cite{Symanzik}, in which the regularized action is improved
through a judicious inclusion of irrelevant operators with increasing
dimensionality. Thus far, most efforts have been directed towards
${\cal O}(a^1)$ improvement; this is automatic in some cases
(i.e. requires no tuning of parameters), by symmetry considerations
alone. Such is the case, for example, of the twisted mass formulation
of QCD~\cite{FGSW,FR} 
at maximal twist, where certain observables are ${\cal O}(a^1)$
improved, as a consequence of symmetries of the fermion action: Setting the
maximal twist requires the tuning of only a single parameter in the
action, i.e. the critical quark mass, and no further improvement of
the operators is required.

In other cases, such as with the clover fermion action,
${\cal O}(a)$ corrections must be also implemented on individual
operators; such corrections take the form of an additional, finite
(non UV-divergent) renormalization or an admixture of appropriate
higher dimensional operators. Determining the values of the
renormalization functions or mixing coefficients requires an
evaluation of appropriate Green's functions, as dictated by the choice
of renormalization scheme; these Green's functions can be evaluated
perturbatively or nonperturbatively.

As regards the perturbative evaluation of Green's functions for the
``ultralocal'' fermion bilinear operators $O^\Gamma_a =
\bar\Psi\lambda_a\Gamma\Psi$ 
($\Gamma$ denotes all possible distinct products of Dirac matrices,
and $\lambda_a$ is a flavor symmetry generator) and the related
fermion propagator, the following types of calculations have appeared
thus far in the literature: $(i)$ One-loop calculations to ${\cal
  O}(a^0, \,\ln a)$ have been performed in the past several years for
a wide variety of actions, ranging from Wilson fermions/gluons to
overlap fermions and Symanzik gluons
\cite{MZ,Aoki98,Capitani,AFPV,CG,HPRSS,IP}. $(ii)$ There exist one-loop
computations of ${\cal O}(a^1)$ corrections, with an arbitrary fermion
mass~\cite{ANTU,Capitani}. $(iii)$ The first two-loop calculations of
Green's functions for $O^\Gamma_a$ were
completed recently, to ${\cal O}(a^0)$, for Wilson/clover/twisted-mass
fermions and Wilson gluons~\cite{SP,SP2}. $(iv)$ A number of 
${\cal O}(a^0)$ results have also been obtained by means of stochastic 
perturbation theory~\cite{Miccio,DiRenzo,Scorzato}.

One-loop computations of ${\cal O}(a^2)$ corrections did not exist to
date; indeed they present some novel difficulties, as compared to
${\cal O}(a^1)$. Extending ${\cal O}(a^0)$ calculations up to ${\cal O}(a^1)$ does not
bring in any novel types of singularities: For instance, terms which
were convergent to ${\cal O}(a^0)$ may now develop at worst an infrared (IR)
logarithmic singularity in 4 dimensions and the way to treat
such singularities is well known; also, in most of the cases, e.g. for
$m=0$, terms which were already IR divergent to ${\cal O}(a^0)$
will not contribute to ${\cal O}(a^1)$, by parity of loop integration. 
On the contrary, many IR singularities encountered at ${\cal
  O}(a^2)$ would persist even up to 6 dimensions, making their extraction
more delicate. In addition to that, there appear Lorentz non-invariant
contributions in ${\cal O}(a^2)$ terms, such as $\sum_\mu p_\mu^4/p^2$
($p$: external momentum).

In this paper we present a one-loop perturbative calculation, to
${\cal O}(a^2)$, of the quantum 
corrections to the fermion propagator and to the complete basis of local
fermion bilinear currents $\bar\Psi\Gamma\Psi$, using massless
fermions described by the Wilson/clover action. We use a 3-parameter
family of Symanzik improved gluon actions, comprising all cases
which are in common use (Wilson, tree-level Symanzik,
Iwasaki~\cite{Iwasaki}, DBW2~\cite{DBW2}, L\"uscher-Weisz~\cite{LW1,LW2}).
All calculations have been performed for generic
values of the gauge parameter. Also, by virtue of working in a massless
scheme, all of our results are applicable to other ultralocal
fermion actions as well, such as the twisted mass or Osterwalder-Seiler
action~\cite{OS}. Our results can be used to
construct ${\cal O}(a^2)$ improved definitions of the fermion
bilinears. In particular, they will be used in Ref.~\cite{ETMC} to
improve the nonperturbative determinations, with the RI-MOM method~\cite{RIMOM}, 
of renormalization constants of bilinear quark operators.

This paper is organized as follows: Section II is an outline of our
calculational procedure; Section III describes in detail the
evaluation of a prototype IR divergent integral; Sections IV and V present
the corrections to the propagator and to fermion bilinears,
respectively; Section VI contains a discussion and concluding
remarks. Appendix A contains a basis of the
divergent integrals which appear in the calculation, evaluated to the
required order in $a$.
\newpage

\section{Description of the calculation}
\label{method}
Our calculation makes use of the clover (SW) action for fermions; for
$N_f$ flavor species this action reads, in standard notation,
\bea
S_F &=& \sum_{f}\sum_{x} (4r+m_f)\bar{\psi}_{f}(x)\psi_f(x)\nonumber \\
&-& {1\over 2}\sum_{f}\sum_{x,\,\mu}\bigg{[}\bar{\psi}_{f}(x) \left( r - \gamma_\mu\right)
U_{x,\, x+\mu}\psi_f(x+{\mu}) 
+\bar{\psi}_f(x+{\mu})\left( r + \gamma_\mu\right)U_{x+\mu,\,x}\psi_{f}(x)\bigg{]}\nonumber \\
&-& {1\over 4}\,c_{\rm SW}\,\sum_{f}\sum_{x,\,\mu,\,\nu} \bar{\psi}_{f}(x)
\sigma_{\mu\nu} {\hat F}_{\mu\nu}(x) \psi_f(x),
\label{clover}
\eea
The Wilson parameter $r$ is set to $r=1$; $f$ is a flavor
index; $\sigma_{\mu\nu} =[\gamma_\mu,\,\gamma_\nu]/2$\,; the clover
coefficient $c_{\rm SW}$ is kept as a free parameter throughout. Powers of
the lattice spacing $a$ have been omitted and may be directly
reinserted by dimensional counting. The tensor $\hat{F}_{\mu\nu}$
is a lattice representation of the gluon field tensor, defined through
\bea
{\hat F}_{\mu\nu} &\equiv& {1\over{8}}\,(Q_{\mu\nu} - Q_{\nu\mu})
\eea
where $Q_{\mu\nu}$ is the sum of the plaquette loops
\begin{eqnarray}
Q_{\mu\nu} &=& U_{x,\, x+\mu}U_{x+\mu,\, x+\mu+\nu}U_{x+\mu+\nu,\, x+\nu}U_{x+\nu,\, x}
+ U_{ x,\, x+ \nu}U_{ x+ \nu,\, x+ \nu- \mu}U_{ x+ \nu- \mu,\, x- \mu}U_{ x- \mu,\, x} \nonumber \\
&+& U_{ x,\, x- \mu}U_{ x- \mu,\, x- \mu- \nu}U_{ x- \mu- \nu,\, x- \nu}U_{ x- \nu,\, x}
+ U_{ x,\, x- \nu}U_{ x- \nu,\, x- \nu+ \mu}U_{ x- \nu+ \mu,\, x+ \mu}U_{ x+ \mu,\, x}
\end{eqnarray}

We perform our calculation for mass independent renormalization
schemes, so that $m_f=0$; this simplifies the
algebraic expressions, but at the same time requires special treatment
when it comes to IR singularities. By taking $m_f=0$, our
calculation and results are identical also for the twisted mass
action and the Osterwalder-Seiler action in the chiral limit (in the so called 
twisted mass basis).

\bigskip
For gluons we employ the Symanzik improved action, involving
Wilson loops with 4 and 6 links\footnote{$1\times 1$ {\em plaquette},
  $1\times 2$ {\em rectangle}, $1\times 2$ {\em chair} (bent
  rectangle), and $1\times 1\times 1$ {\em parallelogram} wrapped
  around an elementary 3-d cube.}, which is given by the relation
\begin{eqnarray}
\hspace{-1cm}
S_G&=&\frac{2}{g_0^2} \Bigl[ c_0 \sum_{\rm plaquette} {\rm Re\,Tr\,}\{1-U_{\rm plaquette}\} 
+ c_1 \sum_{\rm rectangle} {\rm Re \, Tr\,}\{1- U_{\rm rectangle}\} \nonumber \\
\hspace{-1cm}
&&\phantom{\frac{2}{g_0^2}} +c_2 \sum_{\rm chair} {\rm Re\, Tr\,}\{1-U_{\rm chair}\}
+c_3 \sum_{\rm parallelogram} {\rm Re \,Tr\,}\{1-U_{\rm parallelogram}\}\Bigr]
\label{Symanzik}
\end{eqnarray}
The coefficients $c_i$ can in principle be chosen arbitrarily, subject
to the following normalization condition, which ensures the correct
classical continuum limit of the action
\begin{equation}
c_0 + 8 c_1 + 16 c_2 + 8 c_3 = 1
\label{norm}
\end{equation}
Some popular choices of values for $c_i$ used in numerical simulations
will be considered in this work, and are itemized in Table~\ref{tab1};
they are normally tuned in a way as to ensure ${\cal O}(a^2)$ improvement in the 
pure gluon sector.
Our one-loop Feynman diagrams do not
involve pure gluon vertices, and the gluon propagator depends only
on three combinations of the Symanzik parameters: $C_0 \equiv 
c_0 + 8 c_1 + 16 c_2 + 8 c_3 \,(=1)$, $C_1 \equiv c_2 + c_3$, 
$C_2 \equiv c_1 - c_2 - c_3$\,; therefore, with no loss of generality all
these sets of values have $c_2=0$.

\bigskip
For the algebraic operations involved in evaluating the Feynman
diagrams relevant to this calculation, we make use of our symbolic
package in Mathematica. Next, we briefly describe the required steps: 

\vskip 0.35cm \noindent
{$\bullet\,$\underline{Algebraic manipulations:}}\\
The first step in evaluating each diagram is the contraction among
vertices, which is performed automatically once the algebraic
expression of the vertices and the
topology (``incidence matrix'') of the diagram are specified. The
outcome of the contraction 
is a preliminary expression for the diagram; there follow
simplifications of the color dependence, Dirac matrices and
tensor structures. We also fully exploit symmetries of the theory
(periodicity, reflection, conjugation, hypercubic, etc.) to limit the
proliferation of the algebraic expressions.

\vskip 0.35cm\noindent
{$\bullet\,$\underline{Dependence on external momentum:}}\\
Even though one-loop computations are normally a straighforward
procedure, extending to ${\cal O}(a^2)$ introduces several
complications, especially when isolating
logarithms and Lorentz non-invariant terms. As a first task we
want to  reduce the number of infrared divergent integrals to a
minimal set. To do this, we use two kinds of subtractions among the
propagators, using the simple equalities 
\bea
\frac{1}{\tilde q^2}&=&\frac{1}{\hat{q}^2}+
\Bigg{\{}\frac{4\sum_\mu \sin^4(q_\mu/2)-4\left(\sum_\mu \sin^2(q_\mu/2)\right)^2}
{\tilde q^2 \,\hat{q}^2}\Bigg{\}}\label{sub1}\\[3ex]
D(q)&=& D_{plaq}(q) + \Big{\{}D(q) - D_{plaq}(q) \Big{\}} \nonumber \\[1.5ex]
&=&  D_{plaq}(q) + D_{plaq}(q) \Big{\{}D^{-1}_{plaq}(q) - D^{-1}(q) \Big{\}} D(q)
\label{sub2}
\eea
where $q$ stands for $k$ or $k+a\,p$, and $k$ ($p$) is the loop (external)
momentum. The denominator of the fermion propagator, $\tilde q^2$, is
defined as 
\be
\tilde q^2 = \sum_\mu \sin^2(q_\mu)+
\left(m_f+\frac{r}{2} \hat{q}^2\right)^2,\quad
\hat{q}^2 =4\sum_\mu \sin^2(\frac{q_\mu}{2})
\ee
For the present work, one sets $m_f=0$ and $r=1$, as used
in Eq.~(\ref{sub1});
$D$ is the $4\times4$ Symanzik gluon propagator; the
expression for the matrix $\left(D^{-1}_{plaq}(q) - D^{-1}(q)
\right)$, which is ${\cal O}(q^4)$, is independent of the gauge
parameter, $\lambda$, and it can be easily obtained in closed
form. Moreover, we have
\be
\bigl(D_{plaq}(q)\bigr)_{\mu\nu} = \frac{\delta_{\mu\nu}}{\hat{q}^2} -
(1-\lambda)\frac{\hat{q}_\mu\,\hat{q}_\nu}{(\hat{q}^2)^2}
\ee
Terms in curly brackets of Eqs.~(\ref{sub1}) and (\ref{sub2}) are less
IR divergent than their unsubtracted counterparts, by 
two powers in the momentum. These subtractions are performed
iteratively until all primitively divergent integrals (initially
depending on the fermion and the Symanzik propagator) are expressed in
terms of the Wilson gluon propagator. 

Having reduced the number of distinct divergent integrals down to a
minimum, the most laborious task is the computation of these integrals,
which is performed in a noninteger number of dimensions $D>4$.
Ultraviolet divergences are explicitly isolated \`a la Zimmermann and
evaluated as in the continuum. The remainders are $D$-dimensional,
parameter-free, zero external momentum lattice integrals which can be recast in
terms of Bessel functions, and finally expressed as sums of a pole
part plus numerical constants. We analytically evaluate an extensive
basis of superficially divergent loop integrals, listed in
Eqs. (\ref{int1}) - (\ref{int10}) of Appendix~\ref{appA}; a few of these were
calculated in Ref.~\cite{PV}. The integrals of Eqs.~(\ref{int1}),
(\ref{int2}), (\ref{int5}), are the most demanding ones in the list;
they must be evaluated to two further orders in $a$, beyond the order
at which an IR divergence initially sets in. As a consequence,
their evaluation requires going to $D>6$ dimensions. Fortunately, they are a
sufficient basis for all massless integrals which can appear in any ${\cal
  O}(a^2)$ one-loop calculation; that is, any such computation can be
recast in terms of (\ref{int1}), (\ref{int2}), (\ref{int5}), plus
other integrals which are more readily handled. A correct way to
evaluate (\ref{int1}), (\ref{int2}),~(\ref{int5}) has not been
presented previously in the literature, despite their central role in
${\cal O}(a^2)$ calculations, and this has prevented one-loop computations to
${\cal O}(a^2)$ thus far. The calculation of such an integral is given
in detail in the next section.

\bigskip
Terms which are IR convergent can be treated by Taylor expansion in
$ap$ to the desired order. Alternatively, the extraction of the $a
p$ dependence may be
performed using iteratively subtractions of the form
\be
f(k+a\,p) = f(k) + \Big[f(k+a\,p)-f(k)\Big]
\ee
This leads to exact relations such as the following ones
\bea
\frac{1}{\widetilde{k+a\,p}^2}&=&\frac{1}{\tilde{k}^2}-
\frac{\sum_\mu\sin(2k_\mu+a\,p_\mu)\sin(a\,p_\mu)}{\widetilde{k+a\,p}^2\,\tilde{k}^2}\nonumber\\[1ex]
&-&\frac{\sum_\mu\sin(k_\mu+\frac{a\,p_\mu}{2})\sin(\frac{a\,p_\mu}{2})
\left(\hat{k}^2 + \widehat{k+a\,p}^2\right)}
{\widetilde{k{+}a\,p}^2\,\tilde{k}^2}\\[5ex]
\frac{1}{\widehat{k+a\,p}^2}&=&\frac{1}{\hat{k}^2}
-\frac{4\sum_\mu\sin(k_\mu+\frac{a\,p_\mu}{2})\sin(\frac{a\,p_\mu}{2})}
{\widehat{k+a\,p}^2\,\hat{k}^2}
\eea
In these relations the exact $ap$ dependence of the remainders is
under full control; this type of subtraction is especially useful when
applied to the Symanzik propagator.

\vskip 0.35cm\noindent
{$\bullet\,$\underline{Numerical integration:}}\\
The required numerical integrations of the algebraic expressions for
the loop integrands (a total of $\sim$ 40,000 terms) are performed by
highly optimized Fortran programs; these are generated by our
Mathematica `integrator' routine. Each integral is expressed as a sum
over the discrete Brillouin zone of finite lattices, with varying size
$L$ ($4^4\leq L^4 \leq 128^4$), and evaluated for all values of the
Symanzik coefficients listed in Table~\ref{tab1} (corresponding to the
Plaquette, Symanzik, Iwasaki, TILW and DBW2 action).

\vskip 0.35cm\noindent
{$\bullet\,$\underline{Extrapolation:}}\\
The last part of the evaluation is the extrapolation of the numerical
results to infinite lattice size. This procedure entails a systematic
error, which is reliably estimated, using a sophisticated inference
technique; for one-loop quantities we expect a fractional error smaller
than $10^{-7}$.

\section{Evaluation of a primitively divergent integral}
\label{appB}
Divergent integrals which appear in calculations up to ${\cal
  O}(a^1)$ may be evaluated using the standard procedure of Kawai et
al.~\cite{KNS}, in which one subtracts and adds to the
original integrand its naive Taylor expansion, to the appropriate
order with respect to $a$, in $D\to 4^+$ dimensions: The subtracted
integrand, being UV convergent, is calculated in the continuum
limit $a\to 0$, using the methods of Ref.~\cite{CT}, while the Taylor
expansion terms are recast in terms of Bessel functions and are
evaluated in the limit $\epsilon\to 0$ ($\epsilon \equiv (4-D)/2$).

In contrast to the above, some of the integrals in the present work,
given that they 
must be evaluated to ${\cal O}(a^2)$, have Taylor expansions which
remain IR divergent all the way up to $D\leq 6$
dimensions. A related difficulty regards Kawai's procedure:
Subtracting from the original integral its Taylor expansion in
$D$-dimensions to the appropriate order, the UV-convergent subtracted
expression at which one
arrives can no longer be evaluated in the continuum limit by naively
setting $a \to 0$, because there will be ${\cal O}(a^2)$ corrections
which must not be neglected. 
These novel difficulties plague integrals \ref{int1}, \ref{int2},
\ref{int5}, of Appendix~\ref{appA}. Using a combination of momentum
shifts, integration by parts and trigonometric identities, one may
express \ref{int2} and \ref{int5} in terms of \ref{int1} and other less divergent
integrals. Thus, it suffices to address the evaluation of \ref{int1}
\be
A1(p) = \int_{-\pi}^\pi \frac{d^4k}{(2\pi)^4}\frac{1}{\hat k^2\, \widehat{k+a\,p}^2} 
\ee
This is a prototype case of an integral which is IR divergent in
$D\leq 6$ dimensions; in fact, all other integrals encountered in the
present calculation may be expressed in terms of $A1(p)$ plus other
integrals which are IR convergent at $D>4$ (and are thus amenable to a
more standard treatment).

First we split the original integrand $I$ into two parts
\be
I\equiv \frac{1}{\hat k^2\, \widehat{k+a\,p}^2} =  I_1 + I_2
\ee
where $I_2$ is obtained from $I$ by a series expansion, with respect to
the arguments of all trigonometric functions, to subleading order;
$I_1$ is simply the remainder $I-I_2$
\bea
I_1 &=& \frac{k^2-\frac{k^4}{12}-\hat{k}^2}{k^2\,\hat{k}^2\,\widehat{k+a\,p}^2}
+\frac{k^4\left(k^2-\hat{k}^2\right)}{12\,\left(k^2\right)^2\,\hat{k}^2\,\widehat{k+a\,p}^2}
+\frac{k^4\left((k+a\,p)^2-\widehat{k+a\,p}^2\right)}{12\,\left(k^2\right)^2\,(k+a\,p)^2\,\widehat{k+a\,p}^2}
\nonumber \\[2.5ex]
&+&\frac{(k+a\,p)^2-\frac{(k+a\,p)^4}{12}-\widehat{k+a\,p}^2}{k^2\,(k+a\,p)^2\,\widehat{k+a\,p}^2}
+\frac{(k+a\,p)^4\left((k+a\,p)^2-\widehat{k+a\,p}^2\right)}
{12\,k^2\,\left((k+a\,p)^2\right)^2\,\widehat{k+a\,p}^2}\\[3.5ex]
I_2 &=& \frac{1}{k^2\,(k+a\,p)^2}
+\left[\frac{(k+a\,p)^4}{12\,k^2\,\left((k+a\,p)^2\right)^2}
+ \frac{k^4}{12\,\left(k^2\right)^2\,(k+a\,p)^2} \right]
\eea
($q^4 \equiv \sum_\mu q_\mu^4$). $I_2$ is free of trigonometric
functions, while $I_1$ is naively Taylor expandable to
${\cal O}(a^2)$; its integral equals 
\be
\int_{-\pi}^\pi \frac{d^4k}{(2\pi)^4} I_1 = 0.004210419649(1) +
a^2\,p^2\, 0.0002770631001(3) + {\cal O}(a^4, a^4 \ln a)
\ee
The errors appearing in the above equation come from extrapolations to
infinite lattice size.

To evaluate the integral of $I_2$ we split the hypercubic integration
region into a sphere of arbitrary radius $\mu$ about the origin ($\mu
\leq \pi$) plus the rest
\be
\int_{-\pi}^\pi = \int_{|k|\leq\mu} +  
\left(\int_{-\pi}^\pi - \int_{|k|\leq\mu}  \right)
\ee
The integral outside the sphere is free of IR divergences and is
thus Taylor expandable to any order, giving\footnote{Due to its
  peculiar domain, this integral has been evaluated by a Monte Carlo
  routine, rather than as a sum over lattice points. The errors in
  Eq.~(\ref{I2}) are thus Monte Carlo errors.} (for $\mu = 3.14155$)
\be
\left(\int_{-\pi}^\pi - \int_{|k|\leq\mu}\right) 
\frac{d^4k}{(2\pi)^4} I_2 = 6.42919(3)\,10^{-3} + a^2\,p^2\,6.2034(1)\,10^{-5}
+ {\cal O}(a^4)
\label{I2}
\ee
We are now left with the integral of $I_2$ over a sphere. The most
infrared divergent part of $I_2$ is $1/(k^2\,(k + a\,p)^2)$, with IR 
degree of divergence -4, and can be integrated {\it exactly},
giving
\be
\int_{|k|\leq\mu} \frac{d^4k}{(2\pi)^4} \frac{1}{k^2\,(k + a\,p)^2}=
 \frac{1}{16\pi^2}\left(1-\ln(\frac{a^2\,p^2}{\mu^2})\right)
\ee
The remaining two terms comprising $I_2$ have IR degree of divergence
-2, thus their calculation
to ${\cal O}(a^2)$ can be performed in $D$-dimensions, with $D$
slightly greater that 4. Let us illustrate the procedure with one of
these terms: $k^4/(\left(k^2\right)^2\,(k+a\,p)^4)$. By
appropriate substitutions of
\be
\frac{1}{(k+\bar{p})^2} = \frac{1}{k^2} + \frac{-2(k \cdot
  \bar{p})-\bar{p}^2}{k^2\,(k+\bar{p})^2}
\qquad (\bar{p} \equiv a\,p)
\ee
we split this term as follows
\bea
\frac{k^4}{\left(k^2\right)^2\,(k+\bar{p})^2} &=& 
\left[\frac{k^4}{\left(k^2\right)^3} + \frac{k^4\,(-2(k \cdot \bar{p})-\bar{p}^2)}{\left(k^2\right)^4} 
+\frac{4\,k^4(k \cdot \bar{p})^2}{\left(k^2\right)^5} \right]\nonumber\\[2ex]
&+&\left(\frac{k^4\left(4(k \cdot \bar{p})\bar{p}^2+(\bar{p}^2)^2 \right)}{\left(k^2\right)^4\,(k+\bar{p})^2}
+\frac{4\,k^4(k \cdot \bar{p})^2 \left(-2(k \cdot \bar{p}) - \bar{p}^2 \right)}{\left(k^2\right)^5\,(k+\bar{p})^2}\right)
\eea
The part in square brackets is polynomial in $a$ and can be integrated easily, using
$D$-dimensional spherical coordinates. The remaining part is
UV-convergent; thus the integration domain can now be recast in the
form
\be
\int_{|k|\leq\mu} = \int_{|k|<\infty} - \int_{\mu\leq|k|<\infty}
\ee
The integral over the whole space can be performed using the methods
of Ref.~\cite{CT}, whereas the integral outside the sphere of radius
$\mu$ is ${\cal O}(a^3)$ and may be safely dropped. The same procedure
is applied to the last term of $I_2$. Adding the contributions from
all the steps described above, we check that the result is independent
of $\mu$.

\section{Correction to the fermion propagator}
\label{prop}

The fermion propagator is the most common example of an off-shell
quantity suffering from ${\cal O}(a)$ effects. Capitani et al. 
\cite{Capitani} have calculated the first order terms in the lattice
spacing for massive fermions. We carried out this calculation beyond
the first order correction, taking into account all terms up to ${\cal
  O}(a^2)$. Our results, to ${\cal O}(a^1)$, are in perfect agreement
with those of Ref.~\cite{Capitani}. The clover coefficient $c_{\rm SW}$
has been considered to be a free parameter and our results are given
as a polynomial of $c_{\rm SW}$. Moreover, the dependence on the
number of colors $N$, the coupling constant $g$ and the gauge fixing
parameter $\lambda$, is shown explicitly. The Symanzik coefficients,
$c_i$, appear in a nontrivial way in the propagator and, thus, we
tabulate these results for different choices of $c_i$.

The one-loop Feynman diagrams that enter our 2-point Green's function
calculation, are illustrated in Fig. 1. 

\vskip 0.2cm
\begin{center}
\psfig{figure=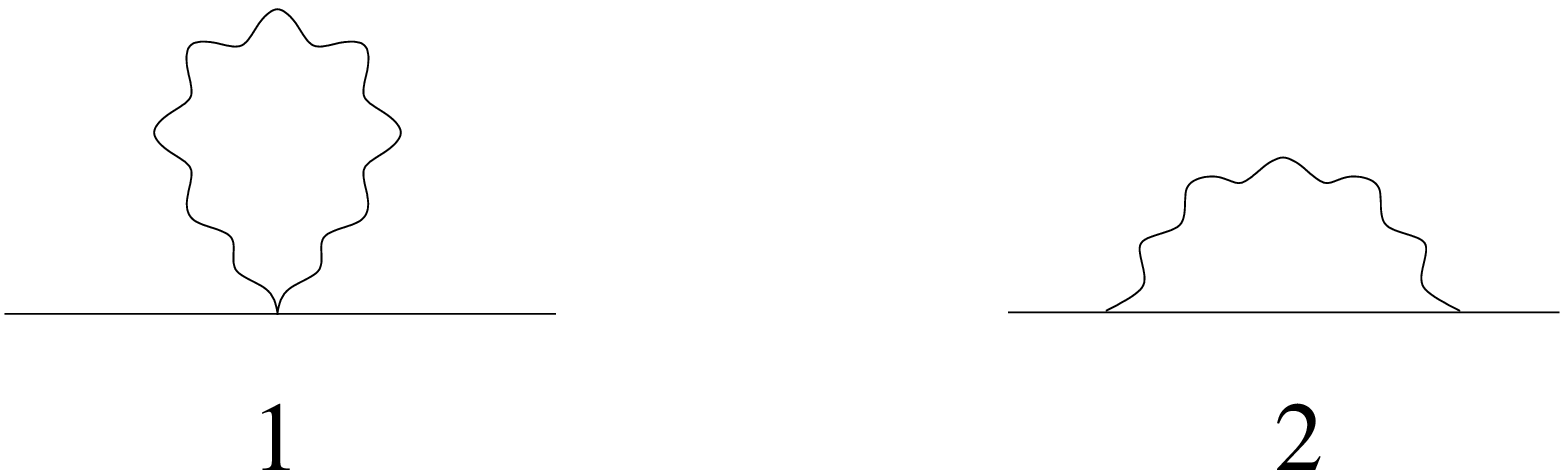,height=2.25truecm}
\end{center}
{\small 
\begin{center}
\begin{minipage}{14cm}
{\bf Fig. 1:} One-loop diagrams contributing to the 
fermion propagator. Wavy (solid) lines represent gluons (fermions).
\end{minipage}
\end{center}}
\vspace{0.75cm}

Next, we provide the total expression for the inverse fermion propagator
$S^{-1}$ as a function of $g,\,N,\,c_{\rm SW},\,\lambda$. Here we
should point out that for dimensional reasons, there is a global
prefactor $1/a$ multiplying our expressions for the inverse
propagator, and thus, the ${\cal O}(a^2)$ correction is achieved by
considering all terms up to ${\cal O}(a^3 p^3)$.  
\bea
\hspace{-0.85cm}
S^{-1}_{(p)}  &=& i \, \pslash+ \frac{a}{2} p^2 - i \frac{a^2}{6}\,\pslash^3\nonumber \\[1.5ex]
&-& i\,\pslash \, \tilde{g}^2 \Big[ \,
      \ve^{(0,1)} -4.79200956(5)\,\lambda  +\ve^{(0,2)}\csw +\ve^{(0,3)}\csw^2
   + \lambda \ln(a^2 p^2) \Big]  \nonumber \\[1.5ex]
&-& a\, p^2\, \tilde{g}^2 \Big[
     \ve^{(1,1)} -3.86388443(2)\,\lambda  +\ve^{(1,2)}\csw +\ve^{(1,3)}\csw^2
 - \frac{1}{2} \left( 3 - 2 \,\lambda  -  3 \csw \right) \, \ln(a^2 p^2)
        \Big] \nonumber \\ [1.5ex]
&-& i\,a^2\, \pslash^3 \, \tilde{g}^2 \Big[\,
\ve^{(2,1)} +0.507001567(9)\,\lambda +\ve^{(2,2)}\csw + \ve^{(2,3)}\csw^2\nonumber \\[1.5ex]
&&\phantom{i\,a^2\, \pslash^3 \, \tilde{g}^2 \Big[}
+\left(\frac{101}{120}-\frac{11}{30}C_2
-\frac{\lambda}{6} \right)\,\ln(a^2 p^2)\Big]\nonumber \\[1.5ex]
&-& i\,a^2\, p^2\,\pslash \, \tilde{g}^2 \Big[\,
\ve^{(2,4)} +1.51604667(9)\,\lambda +\ve^{(2,5)}\csw + \ve^{(2,6)}\csw^2 \nonumber\\[1.5ex]
&&\phantom{i\,a^2\, p^2\,\pslash \, \tilde{g}^2}
+\left(\frac{59}{240}+\frac{c_1}{2}+\frac{C_2}{60}
-\frac{1}{4}\left(\frac{3}{2}\lambda+
\csw+\csw^2\right)\right) \, \ln(a^2 p^2)\Big] \nonumber \\ [1.5ex]
&-& i\,a^2\,\pslash\, \frac{\sum_\mu p^4_\mu}{p^2}\,\tilde{g}^2\Big[\,
-\frac{3}{80}-\frac{C_2}{10} -\frac{5}{48}\lambda \Big] 
\label{propagator}
\eea
where $\tilde{g}^2 \equiv g^2C_F/(16\pi^2)$, $C_F=(N^2-1)/(2N)$, $C_2=c_1 - c_2 - c_3$, $\pslash^3=
\sum_\mu \gamma_\mu p_\mu^3$, and the specific values $\lambda=1\,(0)$ correspond
to the Feynman (Landau) gauge. The quantities
$\ve^{(i,j)}$ appearing in our results for $S^{-1}$ are numerical
coefficients depending on the Symanzik parameters, calculated for each
action we have considered and tabulated in Tables~\ref{tab2} -
\ref{tab5}; the index $i$ denotes the power of the 
lattice spacing $a$ that they multiply. In all Tables, the
systematic errors in parentheses come from the extrapolation over
finite lattice size $L \to \infty$.

Terms proportional to $1/a$ have been left out of
Eq.~(\ref{propagator}) for conciseness; such terms represent ${\cal O}(g^2)$
corrections to the critical value of the fermion mass.

We observe that the ${\cal O}(a^1)$ logarithms as well as all terms
multiplied by $\lambda$, are independent of the Symanzik coefficients;
on the contrary ${\cal O}(a^2)$ logarithms have a mild dependence on
the Symanzik parameters. A number of Lorentz non-invariant tensors ($\sum_\mu
p^4_\mu,\,\pslash^3$) appear in ${\cal O}(a^2)$ correction terms,
compatibly with hypercubic invariance. 
Finally, our ${\cal O}(a^1)$ results for the Plaquette
action, are in agreement with Eq.~(37) of Ref.~\cite{Capitani}.

To enable cross-checks and comparisons, the per-diagram contributions
$d_1(p),\,d_2(p)$ are presented below. The tadpole diagram 1 of Fig. 1 is free of
logarithmic terms and independent of $c_{\rm SW}$; its final
expression is
\bea
\frac{d_1(p)}{\tilde{g}^2} &=&i\,\pslash\Bigl[\,\tilde\ve^{(0,1)}_1 +3.050262540200(1)\,\lambda\Bigr] +
a\,p^2\Bigl[\,\tilde\ve^{(1,1)}_1 + 1.529131270100(1)\,\lambda\Bigr]\nonumber\\ [1.5ex]
&+&i\,a^2\,\pslash^3\Bigl[\,\tilde\ve^{(2,1)}_1 - 0.509710423367(1)\,\lambda\Bigr]
\label{prop_d1}
\eea
where the numerical values for the Symanzik dependent coefficients
$\tilde\ve^{(i,1)}_1$ are listed in Table~\ref{tab7}. The main
contribution to the propagator correction comes from diagram 2, as
can be seen from the following expression, with $\tilde\ve^{(i,1)}_2$
provided in Table~\ref{tab8}. The remaining terms with coefficients
$\ve^{(i,j)}$ are the same as in Eq.~(\ref{propagator}).
\bea
\frac{d_2(p)}{\tilde{g}^2} &=&i\,\pslash\Bigl[\,\tilde\ve^{(0,1)}_2 -7.850272109(6)\,\lambda  
+\ve^{(0,2)}\csw +\ve^{(0,3)}\csw^2
   + \lambda \ln(a^2 p^2) \Bigr] \nonumber \\ [1.5ex]
&+& a\, p^2\Bigl[\,\tilde\ve^{(1,1)}_2 -5.39301570(2)\,\lambda  +\ve^{(1,2)}\csw +\ve^{(1,3)}\csw^2
 - \frac{1}{2} \left( 3 - 2 \,\lambda  -  3 \csw \right) \, \ln(a^2 p^2)\Bigr] \nonumber \\ [1.5ex] 
&+& i\,a^2\, \pslash^3 \Bigl[\,\tilde\ve^{(2,1)}_2 +1.016711991(9)\,\lambda +\ve^{(2,2)}\csw 
+ \ve^{(2,3)}\csw^2\nonumber \\ [1.5ex] 
&&\phantom{i\,a^2\, \pslash^3 \Bigl[}
+\left(\frac{101}{120}-\frac{11}{30}C_2
-\frac{\lambda}{6} \right)\,\ln(a^2 p^2)\Bigr]\nonumber \\ [1.5ex]
&+& i\,a^2\, p^2\,\pslash \Bigl[\ve^{(2,4)} +1.51604667(9)\,\lambda 
+\ve^{(2,5)}\csw + \ve^{(2,6)}\csw^2 \nonumber\\ [1.5ex]
&&\phantom{i\,a^2\, p^2\,\pslash \Bigl[}
+\left(\frac{59}{240}+\frac{c_1}{2}+\frac{C_2}{60}
-\frac{1}{4}\left(\frac{3}{2}\lambda+
\csw+\csw^2\right)\right) \, \ln(a^2 p^2)\Bigr] \nonumber \\ [1.5ex] 
&+& i\,a^2\,\pslash\, \frac{\sum_\mu p^4_\mu}{p^2}\Bigl[
-\frac{3}{80}-\frac{C_2}{10} -\frac{5}{48}\lambda \Bigr]
\label{prop_d2}
\eea
\vskip 0.25cm
Using our results for the fermion propagator, we can compute the
multiplicative renormalization function of the quark field ($Z_\Psi$).

\section{Fermion bilinear operators}
\label{oper}

In the context of this work we also study the ${\cal O}(a^2)$
corrections to Green's functions of local fermion operators that have the form
$\bar\Psi\Gamma\Psi$. $\Gamma$ corresponds to the following set of
products of the Dirac matrices
\be
\Gamma = \openone,\, \gamma^5,\, \gamma_\mu,\, \gamma^5\gamma_\mu,\,
\gamma^5\sigma_{\mu\nu},\qquad \sigma_{\mu\nu}=\frac{1}{2}[\gamma_\mu,\gamma_\nu]
\label{Gamma}
\ee
for the scalar ($O^S$), pseudoscalar ($O^P$), vector
($O^V$), axial ($O^A$) and tensor ($O^T$)
operator, respectively. We restrict ourselves to forward matrix
elements (2-point Green's functions, zero momentum operator
insertions). We also considered the tensor operator $O^{T'}$,
corresponding to $\Gamma = \sigma_{\mu\nu}$ and checked that the
Green's function coincides with that of $O^T$; this is a
nontrivial check for our calculational procedure.

The only one-particle irreducible Feynman diagram that enters the
calculation of the above operators is shown in Fig. 2.

\begin{center}
\psfig{figure=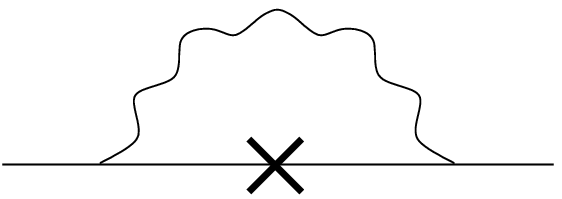,height=1.1truecm}
\end{center}
{\small 
\begin{center}
\begin{minipage}{14cm}
{\bf Fig. 2:} One-loop diagram contributing to the 
bilinear operators. A wavy (solid) line represents gluons (fermions). A
cross denotes the Dirac matrices $\openone$ (scalar), $\gamma^5$
(pseudoscalar), $\gamma_\mu$ (vector), $\gamma^5\gamma_\mu$ (axial),
$\gamma^5\sigma_{\mu\nu}$ (tensor $T$) and $\sigma_{\mu\nu}$ (tensor $T'$).
\end{minipage}
\end{center}}
\vspace{0.75cm}

\bigskip
We show our results for the one-loop
corrections to the amputated 2-point Green's function of each operator
$\bar\Psi\Gamma\Psi$, at momentum $p$
\be
\Lambda^\Gamma(p) = \langle\Psi\,\left(\bar\Psi\Gamma\Psi\right)\,\Bar\Psi\rangle_{(p)}^{amp}
\ee

Our final results are given as a polynomial of $\csw$, in a general
covariant gauge. Since their dependence on the Symanzik parameters,
$c_i$, cannot be written in a closed form, as in the case of the
quark propagator we will tabulate the
numerical coefficients for a variety of choices for $c_i$, in order to
cover a range of values that are used in both perturbative
calculations and numerical simulations.

\medskip
We begin with the ${\cal O}(a^2)$ corrected expression for
$\Lambda^S(p)$; including the tree-level term, we obtain
\noindent
\bea
\hspace{-1cm}
\Lambda^S(p)  &=& 1+ 
\tilde{g}^2\Big[ \,\ve_S^{(0,1)} +5.79200956(5)\,\lambda  +\ve_S^{(0,2)}\csw +\ve_S^{(0,3)}\csw^2
   -\ln(a^2 p^2)\left(3+\lambda\right)\Big]\nonumber \\[1.25ex]
&+&\,a\,i\,\pslash\, \tilde{g}^2 \Big[ \,
\ve_S^{(1,1)} -3.93575928(1)\,\lambda +\ve_S^{(1,2)}\,\csw+\ve_S^{(1,3)}\,\csw^2
+\left(\frac{3}{2}+\lambda +\frac{3}{2}\csw \right)\ln(a^2 p^2)\Big]\nonumber \\[1.25ex]
&+&\,a^2\, p^2 \,  \tilde{g}^2 \Big[\,
\ve_S^{(2,1)} -2.27358943(5)\,\lambda +\ve_S^{(2,2)}\csw + \ve_S^{(2,3)}\csw^2
+\hspace{-0.01cm}\left(-\frac{1}{4}+\frac{3}{4}\lambda +\frac{3}{2}\csw \right)\ln(a^2 p^2)\Big]
\nonumber \\[1.25ex]
&+&\,a^2\,\frac{\sum_\mu p_\mu^4}{p^2}\,  \tilde{g}^2 \Big[\,
\frac{13}{24}+\frac{C_2}{2} -\frac{\lambda}{8} \Bigr]
\label{scalar}
\eea
The numerical coefficients $\ve_S^{(0,i)}$, $\ve_S^{(1,i)}$
and $\ve_S^{(2,i)}$ with their systematic errors are presented in
Tables~\ref{tab9},~\ref{tab10} and \ref{tab11}, respectively.

One might attempt to use the ${\cal O}(a)$ corrections computed above
in order to devise an improved operator, with suppressed finite-$a$
artifacts; it should be noted,
however, that improvement by means of local operators, as permitted
by Quantum Field Theory, is not sufficient to warrant a complete
cancellation of ${\cal O}(a^2)$ terms in Green's functions, since the
latter contain also terms with non-polynomial momentum dependence,
such as $\sum_\mu p_\mu^4/p^2$. Thus, at best, one can achieve full 
${\cal O}(a^2)$ improvement only on-shell, or approximate improvement
near a given 
reference momentum scale. Such non-polynomial terms are not
present at ${\cal O}(a^1)$. This comment applies also to the remaining
operators we examine below. 

\bigskip
Next, we turn to $\Lambda^P(p)$, where Symanzik dependent coefficients,
$\ve_P^{(i,j)}$, are tabulated in Table~\ref{tab12}.
The pseudoscalar operator is free of ${\cal O}(a^1)$ terms;
moreover, all contributions linear in $\csw$ vanish
\bea
\Lambda^P(p)  &=& \gamma^5 + \gamma^5\, \tilde{g}^2 \Big[ \,
      \ve_P^{(0,1)} +5.79200956(5)\,\lambda  +\ve_P^{(0,2)}\csw^2
   -\ln(a^2 p^2)\left(3+\lambda\right)\Big]\nonumber \\[1.25ex]
&+&\,a^2\, p^2 \,\gamma^5\, \tilde{g}^2 \Big[\,
\ve_P^{(2,1)} -0.83810121(5)\,\lambda +\ve_P^{(2,2)}\csw^2
+\hspace{-0.01cm}\left(-\frac{1}{4}+\frac{\lambda}{4}\right)\ln(a^2 p^2)\Big]
\nonumber \\[1.25ex]
&+&\,a^2\, \frac{\sum_\mu p_\mu^4}{p^2}\,\gamma^5\, \tilde{g}^2 \Big[\,
\frac{13}{24}+\frac{C_2}{2}-\frac{\lambda}{8} \Bigr]
\label{pseudoscalar}
\eea

\bigskip
The ${\cal O}(a^2)$ corrected expressions for $\Lambda^V(p)$ and $\Lambda^A(p)$
are more complicated, compared to the scalar and pseudoscalar
amputated Green's functions, in the sense that momentum dependence
assumes a variety of functional forms; this fact also introduces
several coefficients which depend on the Symanzik parameters
\bea
\Lambda^V(p) &=& \gamma_\mu 
+\frac{\pslash\,p_\mu}{p^2}\, \tilde{g}^2 \Big[-2\,\lambda \Big]\nonumber \\[1.25ex]
&+& \gamma_\mu \,\tilde{g}^2 \Big[ \,\ve_V^{(0,1)} +4.79200956(5)\,\lambda
+\ve_V^{(0,2)}\csw+\ve_V^{(0,3)}\csw^2 -\lambda\ln(a^2 p^2)\Big]\nonumber \\[1.25ex]
&+&a\,i\,p_\mu \,\tilde{g}^2 \Big[\,\ve_V^{(1,1)} -0.93575928(1)\,\lambda
+\ve_V^{(1,2)}\csw+\ve_V^{(1,3)}\csw^2 \nonumber \\[1.25ex]
&&\phantom{a\,i\,p_\mu \,\tilde{g}^2 \Big[}
+\left(-3+\lambda+3\,\csw
\right)\ln(a^2 p^2) \Big]\nonumber \\[1.25ex] 
&+&a^2\,\gamma_\mu\, p_\mu^2 \,\tilde{g}^2 \Big[\,\ve_V^{(2,1)}+\frac{\lambda}{8}
+\ve_V^{(2,2)}\csw+\ve_V^{(2,3)}\csw^2 
+\left(-\frac{53}{120}+\frac{11}{10}C_2 \right)
\ln(a^2 p^2)\Big]\nonumber \\[1.25ex] 
&+&a^2\,\gamma_\mu\, p^2 \, \tilde{g}^2 \Big[\,\ve_V^{(2,4)} -0.8110353(1)\,\lambda
+\ve_V^{(2,5)}\csw+\ve_V^{(2,6)}\csw^2\nonumber \\[1.25ex] 
&&\phantom{a^2\,\gamma_\mu\, p^2\, \tilde{g}^2 \Big[\,}
+\left(\frac{11}{240}-\frac{c_1}{2}-\frac{C_2}{60}
+\frac{\lambda}{8}-\frac{5}{12}\csw+\frac{\csw^2}{4}\right)\ln(a^2
p^2)\Big]\nonumber\\[1.25ex] 
&+&a^2\,\pslash\, p_\mu\, \tilde{g}^2 \Big[\,\ve_V^{(2,7)} +0.2436436(1)\,\lambda
+\ve_V^{(2,8)}\csw+\ve_V^{(2,9)}\csw^2\nonumber \\[1.25ex] 
&&\phantom{a^2\, \pslash\,p_\mu\, \tilde{g}^2 \Big[\,}
+\left(-\frac{149}{120}-c_1-\frac{C_2}{30}
+\frac{\lambda}{4}+\frac{\csw}{6}+\frac{\csw^2}{2}\right)\ln(a^2
p^2)\Big]\nonumber \\[1.25ex] 
&+&a^2\,\gamma_\mu\,\frac{\sum_\rho p_\rho^4}{p^2}\,\tilde{g}^2 \Big[\,
\frac{3}{80}+\frac{C_2}{10}+\frac{5}{48}\,\lambda \Bigr] 
+a^2\,\frac{\pslash^3\,p_\mu}{p^2}\, \tilde{g}^2 \Big[
-\frac{101}{60}+\frac{11}{15}C_2 +\frac{\lambda}{3} \Bigr] \nonumber \\[1.25ex] 
&+&a^2\,\frac{\pslash\, p_\mu^3}{p^2}\, \tilde{g}^2 \Big[
-\frac{1}{60}+\frac{2}{5}C_2 +\frac{\lambda}{12} \Bigr] 
+a^2\,\frac{\pslash\,p_\mu\,\sum_\rho
p_\rho^4}{(p^2)^2}\, \tilde{g}^2 \Big[
-\frac{3}{40}-\frac{C_2}{5} -\frac{5}{24}\,\lambda \Bigr]
\label{vector}
\eea
The numerical values of $\ve_V^{(i,j)}$ for different Symanzik
choices are given in Tables~\ref{tab13} - \ref{tab18}. 
\bea
\Lambda^A(p)&=& \gamma^5\,\gamma_\mu
+\frac{\gamma^5\,\pslash\,p_\mu}{p^2}\, \, \tilde{g}^2 \Big[-2\,\lambda \Big]\nonumber \\[1.25ex]
&+&\gamma^5\,\gamma_\mu \, \tilde{g}^2 \Big[ \,\ve_A^{(0,1)} +4.79200956(5)\,\lambda
+\ve_A^{(0,2)}\csw+\ve_A^{(0,3)}\csw^2 -\lambda\ln(a^2 p^2)\Big]\nonumber \\[1.25ex]
&+&a\,i\,\gamma^5\,\left(\gamma_\mu\, \pslash-p_\mu\right) \, \tilde{g}^2 
\Big[\,\ve_A^{(1,1)} -2.93575928(1)\,\lambda\nonumber \\[1.25ex]
&&\phantom{a\,i\,\gamma^5\,\left(\gamma_\mu\, \pslash-p_\mu\right)\,\tilde{g}^2 \Big[}
+\ve_A^{(1,2)}\csw+\ve_A^{(1,3)}\csw^2 +\lambda\ln(a^2 p^2) \Big]\nonumber \\[1.25ex] 
&+&a^2\,\gamma^5\,\gamma_\mu\, p_\mu^2 \, \tilde{g}^2 \Big[\,\ve_A^{(2,1)}+\frac{\lambda}{8}
+\ve_A^{(2,2)}\csw+\ve_A^{(2,3)}\csw^2 +
\left(-\frac{53}{120}+\frac{11}{10}C_2 \right)\ln(a^2 p^2)\Big]\nonumber \\[1.25ex] 
&+&a^2\,\gamma^5\,\gamma_\mu\, p^2 \, \tilde{g}^2 \Big[\,\ve_A^{(2,4)} -1.7465235(1)\,\lambda
+\ve_A^{(2,5)}\csw+\ve_A^{(2,6)}\csw^2\nonumber \\[1.25ex] 
&&\phantom{a^2\,\gamma^5\,\gamma_\mu\, p^2\, \tilde{g}^2 \Big[\,}
+\left(-\frac{109}{240}-\frac{c_1}{2}-\frac{C_2}{60}
+\frac{5}{8}\lambda+\frac{7}{12}\csw-\frac{\csw^2}{4}\right)\ln(a^2
p^2)\Big]\nonumber\\[1.25ex] 
&+&a^2\,\gamma^5\,\pslash\, p_\mu\, \tilde{g}^2 \Big[\,\ve_A^{(2,7)} +1.1146200(1)\,\lambda
+\ve_A^{(2,8)}\csw+\ve_A^{(2,9)}\csw^2\nonumber \\[1.25ex] 
&&\phantom{a^2\,\gamma^5\,\pslash\,p_\mu\, \tilde{g}^2 \Big[\,}
+\left(\frac{91}{120}-c_1-\frac{C_2}{30}
-\frac{3}{4}\lambda-\frac{5}{6}\csw-\frac{\csw^2}{2}\right)\ln(a^2
p^2)\Big]\nonumber \\[1.25ex] 
&+&a^2\,\gamma^5\,\gamma_\mu\,\frac{\sum_\rho p_\rho^4}{p^2}\,\tilde{g}^2 \Big[\,
\frac{3}{80}+\frac{C_2}{10}+\frac{5}{48}\,\lambda \Bigr] 
+a^2\,\gamma^5\frac{\pslash^3\,p_\mu}{p^2}\, \tilde{g}^2 \Big[
-\frac{101}{60}+\frac{11}{15}C_2 +\frac{\lambda}{3} \Bigr] \nonumber \\[1.25ex] 
&+&a^2\,\gamma^5\frac{\pslash\, p_\mu^3}{p^2}\, \tilde{g}^2 \Big[
-\frac{1}{60}+\frac{2}{5}C_2 +\frac{\lambda}{12} \Bigr] 
+a^2\,\gamma^5\frac{\pslash\,p_\mu\,\sum_\rho
p_\rho^4}{(p^2)^2}\, \tilde{g}^2 \Big[
-\frac{3}{40}-\frac{C_2}{5} -\frac{5}{24}\,\lambda \Bigr]
\label{axial}
\eea
Eq.~(\ref{vector}) and Eq.~(\ref{axial}) have many similar terms, among them the coefficients
\be
\ve_A^{(0,2)}=-\ve_V^{(0,2)}, \quad \ve_A^{(0,3)}=-\ve_V^{(0,3)}
\ee
The rest of the coefficients $\ve_A^{(i,j)}$ appear in Tables~\ref{tab19} - \ref{tab22}.

\bigskip
The remaining Green's functions that we computed are those
corresponding to the tensor bilinears ($T=\gamma^5\sigma_{\mu\nu}$,
$T'=\sigma_{\mu\nu}$), which are the most complicated of all the
operators that we studied. Clearly, the Green's functions
$\Lambda^T(p)$ and $\Lambda^{T'}(p)$,
corresponding to $T$ and $T'$, coincide numerically, even though this fact
is not immediately apparent from their algebraic forms. In fact, we
computed both $\Lambda^T(p)$ and $\Lambda^{T'}(p)$ in two distinct
calculations; their numerical coincidence constitutes a rather nontrivial
check of our results. For the reader's convenience, we present below
both tensor Green's functions.
\bea
\hspace{-0.8cm}
\Lambda^T(p) &=& \gamma^5\,\smn+
\gamma^5\,\smn \, \tilde{g}^2 \Big[ \,\ve_T^{(0,1)} +3.79200956(5)\,\lambda
+\ve_T^{(0,2)}\csw+\ve_T^{(0,3)}\csw^2 +\left(1-\lambda\right)\ln(a^2 p^2)\Big]\nonumber \\[1ex]
&+&a\,i\,\gamma^5\frac{\left(\gamma_\nu\, p_\mu- \gamma_\mu\, p_\nu\right)}{2} \, \tilde{g}^2 \Big[\,\ve_T^{(1,1)} +3.87151852(5)\,\lambda
+\ve_T^{(1,2)}\csw+\ve_T^{(1,3)}\csw^2 \nonumber \\[1ex] 
&&\phantom{a\,i\,\gamma^5\frac{\left(\gamma_\nu\, p_\mu- \gamma_\mu\, p_\nu\right)}{2} \, \tilde{g}^2 \Big[}
+\left(3-2\lambda-\csw\right)\ln(a^2 p^2) \Big]\nonumber \\[1ex] 
&+&a^2\,\gamma^5\frac{\left(\gamma_\mu\,\gamma_\nu\, p_\mu^2-\gamma_\nu\,\gamma_\mu\, p_\nu^2\right)}{2} 
\, \tilde{g}^2 \Big[\,\ve_T^{(2,1)}+\frac{\lambda}{4} +\ve_T^{(2,2)}\,\csw 
+\ve_T^{(2,3)}\csw^2 \Big]\nonumber \\[1ex] 
&+&a^2\,\gamma^5\frac{\left(\gamma_\nu\,\pslash\, p_\mu-\gamma_\mu\,\pslash\,p_\nu\right)}{2}\, \tilde{g}^2 \Big[\,\ve_T^{(2,4)} +0.62097643(2)\,\lambda
+\ve_T^{(2,5)}\csw+\ve_T^{(2,6)}\csw^2\nonumber \\[1ex] 
&&\phantom{a^2\,\gamma^5\frac{\left(\gamma_\nu\,\pslash\, p_\mu-\gamma_\mu\,\pslash\,p_\nu\right)}{2}\, \tilde{g}^2 \Big[}
+\left(2-\lambda-\csw\right)\ln(a^2 p^2)\Big]\nonumber\\[1ex] 
&+&a^2\,\gamma^5\,\smn p^2 \, \tilde{g}^2 \Big[\,\ve_T^{(2,7)} -0.7839694(1)\,\lambda
+\ve_T^{(2,8)}\csw+\ve_T^{(2,9)}\csw^2\nonumber \\[1ex] 
&&\phantom{a^2\,\gamma^5\,\smn p^2 \, \tilde{g}^2 \Big[}
+\left(\,\frac{1}{12}-c_1+\frac{C_2}{3}
-\frac{\csw}{2}\right)\ln(a^2 p^2)\Big]\nonumber\\[1.25ex] 
&+&a^2\,\gamma^5\frac{\left(\gamma_\nu\,\pslash\,p_\mu^3-\gamma_\mu\,\pslash\,p_\nu^3\right)}{2\,p^2}\,\tilde{g}^2 
\Big[-\frac{1}{2}+C_2 +\frac{\lambda}{2} \Bigr]
+a^2\,\gamma^5\frac{\left(p_\mu^3\,p_\nu-p_\nu^3\,p_\mu\right)}{2\,p^2}\, \tilde{g}^2 
\Big[\,\frac{17}{3}+\frac{2}{3}C_2\Bigr]\nonumber\\[1.25ex] 
&+&a^2\,\gamma^5\frac{\left(\gamma_\nu\,\pslash^3\,p_\mu-\gamma_\mu\,\pslash^3\,p_\nu\right)}{2\,p^2}
\, \tilde{g}^2 \Big[\,\frac{17}{6}+\frac{C_2}{3}\Bigr] 
+a^2\,\gamma^5\smn\frac{\sum_\rho p_\rho^4}{p^2}
\, \tilde{g}^2 \Big[-\frac{1}{3} +\frac{C_2}{2}+\frac{\lambda}{3} \Bigr]\nonumber \\[1.25ex] 
&+&a^2\,\gamma^5\frac{\left(\gamma_\mu\,\pslash\,p_\mu^2\,p_\nu-
\gamma_\nu\,\pslash\,p_\mu\,p_\nu^2\right)}{2\,p^2}\,\tilde{g}^2 
\Big[-\frac{7}{3}-\frac{4}{3}C_2 -\frac{\lambda}{2} \Bigr] 
\label{tensor}
\eea
The coefficients $\ve_T^{(i,j)}$ are tabulated in Tables~\ref{tab23} - \ref{tab27}. 

\bigskip
\bea
\hspace{-0.8cm}
\Lambda^{T'}(p) &=& \,\smn+
\,\smn \, \tilde{g}^2 \Big[ \,\ve_{T'}^{(0,1)} +3.79200956(5)\,\lambda
+\ve_{T'}^{(0,2)}\csw+\ve_{T'}^{(0,3)}\csw^2 +\left(1-\lambda\right)\ln(a^2 p^2)\Big]\nonumber \\[1ex]
&+&a\,i\,\frac{\left(\gamma_\nu\, p_\mu- \gamma_\mu\, p_\nu\right)+\smn\, \pslash}{2} \, 
\tilde{g}^2 \Big[\,\ve_{T'}^{(1,1)} -3.87151852(5)\,\lambda
+\ve_{T'}^{(1,2)}\csw+\ve_{T'}^{(1,3)}\csw^2 \nonumber \\[1ex] 
&&\phantom{a\,i\,\frac{\left(\gamma_\nu\, p_\mu- \gamma_\mu\, p_\nu\right)+\smn}{2} \, \tilde{g}^2 \Big[}
+\left(-3+2\lambda+\csw\right)\ln(a^2 p^2) \Big]\nonumber \\[1ex] 
&+&a^2\,\frac{\left(\gamma_\mu\,\gamma_\nu\, p_\mu^2-\gamma_\nu\,\gamma_\mu\, p_\nu^2\right)}{2} 
\, \tilde{g}^2 \Big[\,\ve_{T'}^{(2,1)} +\frac{\lambda}{4} +\ve_{T'}^{(2,2)}\csw 
+\ve_{T'}^{(2,3)}\csw^2 \Big]\nonumber \\[1ex] 
&+&a^2\,\frac{\left(\gamma_\nu\,\pslash\, p_\mu-\gamma_\mu\,\pslash\,p_\nu\right)}{2}\, \tilde{g}^2 \Big[\,\ve_{T'}^{(2,4)} -1.12097643(1)\,\lambda
+\ve_{T'}^{(2,5)}\csw+\ve_{T'}^{(2,6)}\csw^2\nonumber \\[1ex] 
&&\phantom{a^2\,\frac{\left(\gamma_\nu\,\pslash\, p_\mu-\gamma_\mu\,\pslash\,p_\nu\right)}{2}\, \tilde{g}^2 \Big[}
+\left(-2+\lambda+\csw\right)\ln(a^2 p^2)\Big]\nonumber\\[1ex] 
&+&a^2\,\smn p^2 \, \tilde{g}^2 \Big[\,\ve_{T'}^{(2,7)} -1.2194576(1)\,\lambda
+\ve_{T'}^{(2,8)}\csw+\ve_{T'}^{(2,9)}\csw^2\nonumber \\[1ex] 
&&\phantom{a^2\,\smn p^2 \, \tilde{g}^2 \Big[}
+\left(-\frac{11}{12}-c_1 +\frac{C_2}{3} +\frac{\lambda}{2}\right)\ln(a^2 p^2)\Big]\nonumber\\[1.25ex] 
&+&a^2\frac{\left(\gamma_\nu\,\pslash\,p_\mu^3-\gamma_\mu\,\pslash\,p_\nu^3\right)}{2\,p^2}\,\tilde{g}^2 
\Big[-\frac{1}{2}+C_2 +\frac{\lambda}{2} \Bigr]
+a^2\frac{\left(p_\mu^3\,p_\nu-p_\nu^3\,p_\mu\right)}{2\,p^2}\, \tilde{g}^2 
\Big[\,\frac{17}{3}+\frac{2}{3}C_2\Bigr]\nonumber\\[1.25ex] 
&+&a^2\frac{\left(\gamma_\nu\,\pslash^3\,p_\mu-\gamma_\mu\,\pslash^3\,p_\nu\right)}{2\,p^2}
\, \tilde{g}^2 \Big[\,\frac{17}{6}+\frac{C_2}{3}\Bigr] 
+a^2\,\smn\frac{\sum_\rho p_\rho^4}{p^2}
\, \tilde{g}^2 \Big[-\frac{1}{3} +\frac{C_2}{2}+\frac{\lambda}{3} \Bigr]\nonumber \\[1.25ex] 
&+&a^2\frac{\left(\gamma_\mu\,\pslash\,p_\mu^2\,p_\nu-
\gamma_\nu\,\pslash\,p_\mu\,p_\nu^2\right)}{2\,p^2}\,\tilde{g}^2 
\Big[-\frac{7}{3}-\frac{4}{3}C_2 -\frac{\lambda}{2} \Bigr] 
\label{tensorprime}
\eea
Several coefficients $\ve_{T'}$ can be written in terms of $\ve_{T}$
(Eqs.~(\ref{TTprime1}) - (\ref{TTprime2})), while the rest are given
in Tables~\ref{tab30} - \ref{tab31} 
\bea
\ve_{T'}^{(0,1)}=\phantom{-}\ve_T^{(0,1)}, \quad \ve_{T'}^{(0,2)}&=&\phantom{-}\ve_T^{(0,2)},\quad
\ve_{T'}^{(0,3)}=\phantom{-}\ve_T^{(0,3)}, 
\label{TTprime1} \\[2ex]
\ve_{T'}^{(1,1)}=-\ve_T^{(1,1)}, \quad \ve_{T'}^{(1,2)}&=&-\ve_T^{(1,2)},\quad
\ve_{T'}^{(1,3)}=-\ve_T^{(1,3)}, \\[2ex]
\ve_{T'}^{(2,2)}=-\ve_T^{(2,2)}, \quad \ve_{T'}^{(2,3)}=-\ve_T^{(2,3)}&,&
\ve_{T'}^{(2,5)}=-\ve_T^{(2,5)}, \quad \ve_{T'}^{(2,6)}=-\ve_T^{(2,6)}
\label{TTprime2}
\eea

\section{Discussion and Conclusions}
\label{conc}

In this paper we have calculated the fermion propagator $S(p)$ and the
Green's functions $\Lambda^\Gamma(p)$ for the fermion bilinear
operators $\bar\Psi\Gamma\Psi$, where $\Gamma$ stands for any product
of Dirac gamma matrices. Our calculations were performed to one loop
in lattice perturbation theory, using the Wilson/clover fermion
action. For gluons we employed a family of Symanzik improved actions,
parameterized by 3 independent ``Symanzik'' coefficients; explicit
results are presented for some of the most commonly used actions in
this family: Wilson, Tree-level Symanzik, Tadpole improved
L\"uscher-Weisz, Iwasaki and DBW2.

Our calculations extend, to a rather large family of fermion/gluon
actions, results which were previously known to ${\cal O}(a^0)$ and
${\cal O}(a^1)$ (modulo $\ln a$). However, the truly novel feature in
our calculations is that they were performed to second order in the
lattice spacing $a$ (${\cal O}(a^2, a^2\ln a)$). This fact introduces
a number of complications, which are not present in lower order
results. In a nutshell, the reason for these complications is as
follows: The extraction of a further power of $a$ from a Feynman
diagram strengthens, by one unit, the superficial degree of infrared
(IR) divergence of the corresponding integrand over loop
momenta. Thus, a priori, in a ${\cal O}(a^1)$ calculation, loop
integrals would be IR convergent only in $D>5$ dimensions; however, as
can be easily deduced by inspection, the most divergent parts of the
integrands are odd functions of the loop momenta, and will thus vanish
upon integration. What is left behind is a less divergent integrand
which is IR convergent in $D>4$, just as in the case of 
${\cal O}(a^0)$ calculations, and can thus be treated by standard
methods, such as those of Ref.~\cite{KNS}. For ${\cal O}(a^2)$
calculations, on the other hand, integrands are IR convergent only at
$D>6$, and their most divergent parts no longer vanish upon
integration; a naive application of the procedure of Ref.~\cite{KNS}
will fail to produce all ${\cal O}(a^2)$ contributions.
The procedure which we propose in this work for handling the above
difficulty is in fact applicable to any order in $a$. In brief, it
recasts the integrands as a sum of two parts: The first part can be
{\it exactly} evaluated as a function of $a$, while the second part is
naively Taylor expandible, as a polynomial to the desired order in
$a$.

Since the propagator and Green's functions are meant to be used in
mass independent renormalization schemes, our results have been
obtained at vanishing fermionic masses; the case of massive fermions
(including non-degenerate flavors and twisted mass terms) will appear
in a forthcoming publication. Nevertheless, even at vanishing masses,
our final expressions are quite lengthy, since they exhibit a rather
nontrivial dependence on the external momentum ($p$), and they are
explicit functions of the number of colors ($N$), gauge parameter
($\lambda$), lattice spacing ($a$), clover coefficient ($\csw$) and
coupling constant ($g$); furthermore, most numerical coefficients in
these expressions depend on the Symanzik parameters of the gluon
action, and we have tabulated them for the actions we have
selected. For convenience, we accompany this paper with an electronic
document, in the form of a Mathematica input file, allowing the reader
to recover immediately numerical values for any choice of input
parameters. 

One possible use of our results is in constructing improved versions
of the operators $O^\Gamma$, with reduced lattice artifacts. In doing
so, however, one must bear in mind that, unlike the ${\cal O}(a^1)$
case, corrections to ${\cal O}(a^2)$ include expressions which are
non-polynomial in the external momentum and, therefore, cannot be
eliminated by introducing admixtures of local operators. Full
improvement can be achieved at best for on-shell matrix elements only.

Starting from $S(p)$ and $\Lambda^\Gamma(p)$, it is straightforward to
write down the renormalization functions $Z_q$ (for the quark field)
and $Z_\Gamma$ (for the operators $O^\Gamma$) in any renormalization
scheme. $Z_q$ and $Z_\Gamma$, as obtained from $S(p)$ and
$\Lambda^\Gamma(p)$, differ from the corresponding expressions
evaluated at ${\cal O}(a^0)$, by lattice artefact, which are functions of 
$(a\mu)$ ($\mu$: renormalization scale), and vanish as $a\to 0$. 
At the nonzero values of $a$
employed in numerical simulations, these factors are quite
important. Ideally, one would prefer a nonperturbative determination
of renormalization functions; while this is often possible, several
sources of error must be dealt with. A very effective way to proceed
is through a combination of perturbative and nonperturbative results.
This procedure is carried out and explained in detail in a follow-up
work~\cite{ETMC}. Briefly stated, nonperturbative data are
``corrected'' by the perturbative expressions for Green's functions,
and then extrapolated towards small $a$.
As a first illustration of this mixed determination, we show
in Fig. 3 nonperturbative data for $Z_q$ and $Z_V$, determined with the RI-MOM 
method of Ref.~\cite{RIMOM}, before and after the perturbative corrections. 
The results are obtained by using the Symanzik tree-level improved gluon
action at $\beta=3.9$ and the $N_f=2$ twisted mass quark action at maximal 
twist, with gauge field configurations and quark propagators generated by the 
ETM Collaboration\footnote{We thank the members of the ETM Collaboration for 
having provided us with the data of Fig. 3 before publication.}.
While up to discretization effects $Z_V$ is a scale independent quantity,
the (continuum) RG dependence of $Z_q$ on the renormalization scale has
been removed from the results shown in Fig. 3 by evolving the
renormalization constant to a fixed reference scale $\mu_0=1/a$ ($\sim 2$ GeV),
using for the anomalous dimension the 4-loop perturbative expression
computed in Ref.~\cite{Chetyrkin:1999pq}. Thus, the residual dependence
of both $Z_q(\mu_0=1/a)$ and $Z_V$ on $a^2\tilde p^2$ observed in Fig. 3 can be
safely interpreted, at large momenta, as a pure discretization effect.
As illustrated in Fig. 3, the corrected data are virtually flat,
allowing for a safer small-$a$ extrapolation.

\vskip 0.2cm
\begin{center}
\psfig{figure=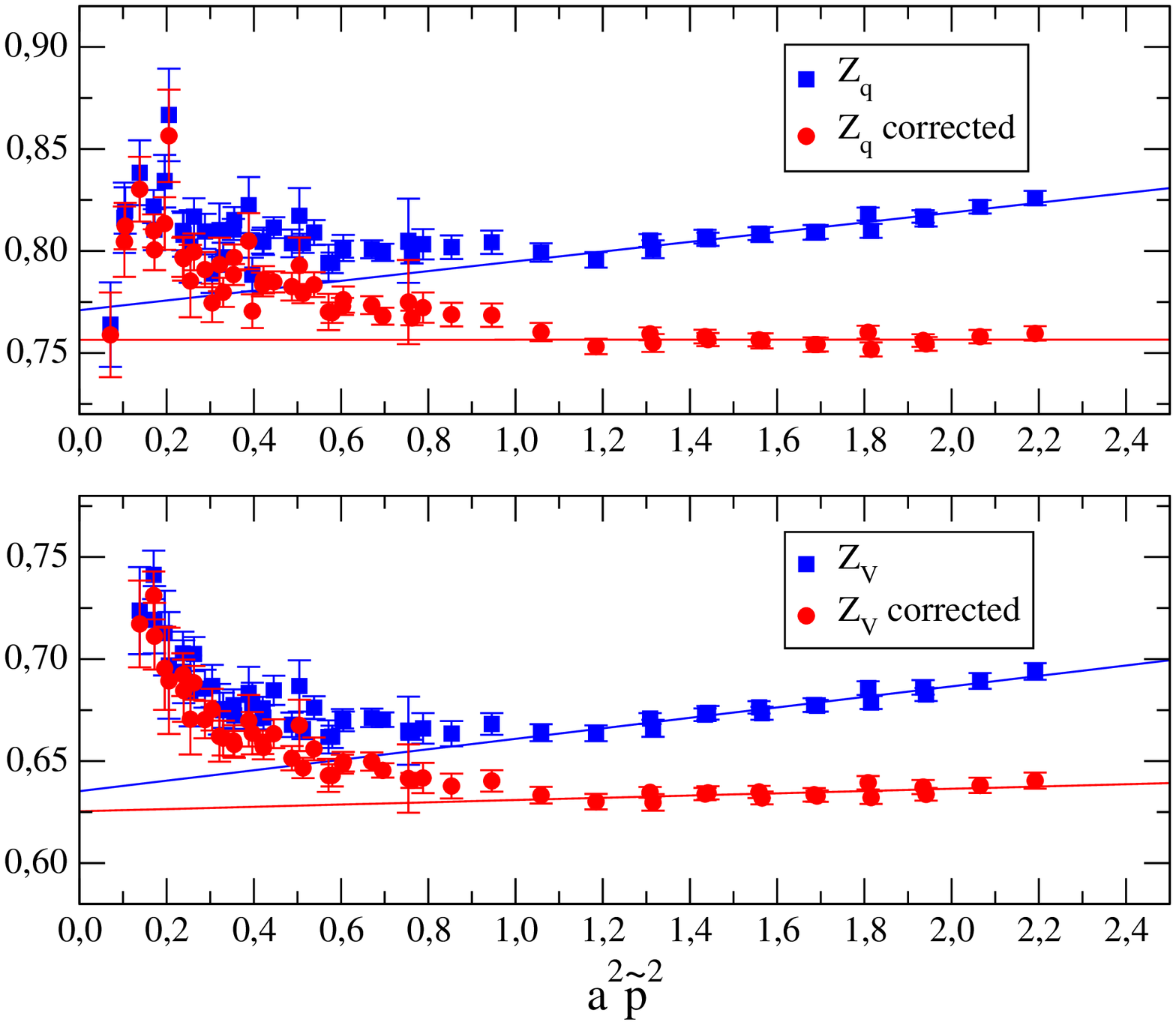,height=22truecm,angle=-90}
\end{center}
{\small 
\begin{center}
\begin{minipage}{14cm}
{\bf Fig. 3:} Non-perturbative data for $Z_q(\mu_0=1/a)$ and $Z_V$, before and
after perturbative corrections. Straight lines are extrapolations to
small $a$. ($a^2 \tilde{p}^2 \equiv \sum_\mu \sin^2(a p_\mu) $)
\end{minipage}
\end{center}}
\vspace{0.75cm}

The techniques employed in this work are readily applicable to the
study of perturbative corrections of other Greens's functions, to any
desired order in $a$. Examples are matrix elements of 4-fermion
operators appearing in effective weak Hamiltonians, and higher
dimension twist-2 fermion bilinears involved in generalized parton
distributions. We will be addressing these issues in forthcoming publications.


\newpage
\appendix
\section{A basis of divergent integrals}
\label{appA}

The most difficult part of this calculation that requires careful
attention is the extraction of the dependence on the external momentum
$p$ and the lattice spacing $a$ from the divergent terms. The
singularities are isolated using the procedure explained in
section~\ref{method}, and here we present the list of primitively
divergent integrals that appeared in our algebraic expressions. 

In the following integrals we define
\bea
\hat{k}_\mu&=&2\,\sin(\frac{k_\mu}{2})\nonumber\\
\hat{k}^2&=&4\sum_\mu \sin^2(\frac{k_\mu}{2})\nonumber\\
\kc_\mu &=&\sin(k_\mu) \nonumber
\eea
In addition, $(\,\,)_S$ means sum over inequivalent permutations. No
summation over the indices $\mu,\,\nu,\,\rho,\,\sigma$ is
implied, unless otherwise stated.
\medskip
\bea
\bullet\int_{-\pi}^\pi \frac{d^4k}{(2\pi)^4}\frac{1}{\hat k^2\, \widehat{k+a\,p}^2} 
&=& 0.036678329075 - \frac{\ln(a^2 p^2)}{16\pi^2}\nonumber\\[0.85ex]
\hspace{-1.25cm}
&+& 0.0000752406(3)\, a^2\,p^2 +a^2\,\frac{\sum_{\mu} p^4_{\mu}}{384\pi^2 \,p^2}+{\cal O}(a^4\,p^4)
\label{int1}\\[2ex]
\nonumber\\
\nonumber\\
\bullet\int_{-\pi}^\pi \frac{d^4k}{(2\pi)^4}\frac{\kc_\mu}{\hat k^2\, \widehat{k+a\,p}^2} 
&=&a\,p_\mu\Bigl[-0.008655827648  +\frac{\ln(a^2 p^2)}{32\pi^2} \nonumber\\[0.85ex]
\hspace{-1.25cm}
&&\phantom{a\,p_\mu\Bigl[}-0.0005107825(2)\,a^2\,p^2 +0.001171329715\,a^2\,p_\mu^2\nonumber\\[0.85ex]
\hspace{-1.25cm}
&&\phantom{a\,p_\mu\Bigl[} -a^2\frac{\sum_{\mu} p^4_{\mu}}{768\pi^2 \,p^2}
 +a^2\, \frac{\ln(a^2 p^2)}{384\pi^2} 
\left(\frac{p^2}{2} -p_\mu^2\right)\Bigr]\nonumber\\[0.85ex]
&&+{\cal O}(a^5\, p^5)
\label{int2}
\eea
\bea
\hspace{-1cm}
\bullet\int_{-\pi}^\pi \frac{d^4k}{(2\pi)^4}\frac{\kc_\mu\,\kc_\nu}{(\hat k^2)^2\, \widehat{k+a\,p}^2} 
&=&\delta_{\mu\nu} \Bigl[\,0.004327913824 - \frac{\ln(a^2 p^2)}{64\pi^2}\nonumber\\[1.5ex]
\hspace{-1cm}
&&\phantom{\delta_{\mu\nu} \Bigl[}
+0.00025539124(8)\,a^2\,p^2 -0.000135654113\,a^2\,p_\mu^2\nonumber\\[1.5ex]
\hspace{-1cm}
&&\phantom{\delta_{\mu\nu} \Bigl[}+a^2\,\frac{\sum_\mu p_\mu^4}{1536\pi^2 \,p^2} 
+a^2\,\frac{\ln(a^2 p^2)}{768\pi^2}\left(p_\mu^2-\frac{p^2}{2} \right)\Bigr]
\nonumber\\[1.5ex]
\hspace{-1cm}
&+&a^2\,p_\mu\,p_\nu \Bigl[\,\frac{1}{32\pi^2\,a^2\,p^2} -0.0003788538(2) 
+\frac{\sum_\mu p_\mu^4}{768\pi^2\,(p^2)^2}\nonumber\\[1.5ex]
\hspace{-1cm}
&&\phantom{a^2\,p_\mu\,p_\nu \Bigl[}
-\frac{(p_\mu^2+p_\nu^2)}{384\pi^2\,p^2}+\frac{\ln(a^2 p^2)}{768\pi^2}
 \Bigr]+{\cal O}(a^4\,p^4)
\label{int5}\\[3ex]
\nonumber\\
\nonumber\\
\hspace{-1.25cm}
\bullet\int_{-\pi}^\pi
\frac{d^4k}{(2\pi)^4}\frac{\kc_\mu\,\kc_\nu}{\hat k^2\,
  \widehat{k+a\,p}^2}
&=& \delta_{\mu\nu} \Bigl[\,0.014966695116 -0.001256484446\,a^2\,p^2 \nonumber\\[0.85ex]
\hspace{-1.25cm}
&&\phantom{\delta_{\mu\nu} \Bigl[}
-0.001027789631\,a^2\,p_\mu^2+ \frac{a^2\,p^2\,\ln(a^2 p^2)}{192\pi^2}\Bigr]\nonumber\\[0.85ex]
\hspace{-1.25cm}
&+&a^2\,p_\mu \,p_\nu \Bigl[0.003970508789 -\frac{\ln(a^2 p^2)}{48\pi^2}\Bigr]
+{\cal O}(a^4\,p^4)
\label{int3}\\[3ex]
\nonumber\\
\nonumber\\
\hspace{-1.25cm}
\bullet\int_{-\pi}^\pi \frac{d^4k}{(2\pi)^4}\frac{(\kc_\mu)^3}{\hat k^2\, \widehat{k+a\,p}^2} 
&=& a\,p_\mu\Bigl[-0.006184131744 +0.001102333439\,a^2\,p^2 \nonumber\\[0.85ex]
\hspace{-1.25cm}
&&\phantom{a\,p_\mu\Bigl[}
-0.000174224479\, a^2\,p_\mu^2 +a^2\,\frac{\ln(a^2p^2)}{64\pi^2}
\left(p_\mu^2-\frac{p^2}{2}\right)\Bigr]\nonumber\\[0.85ex]
&+&{\cal O}(a^5p^5)
\label{int4}\\[3ex]
\nonumber\\
\nonumber\\
\hspace{-1cm}
\bullet\int_{-\pi}^\pi \frac{d^4k}{(2\pi)^4}\frac{\kc_\mu\,\kc_\nu \,\kc_\rho}
{(\hat k^2)^2\, \widehat{k+a\,p}^2} 
&=& 
(\delta_{\nu\rho}\,a\, p_\mu)_S\Bigl[-0.000728769948+\frac{\ln(a^2 p^2)}{192\pi^2}\Bigr]\nonumber\\[1.5ex]
\hspace{-1cm}
&+&0.001027789631 \delta_{\mu\nu\rho}\,a\,p_\mu -a\,\frac{p_\mu\,p_\nu\,p_\rho}{48\pi^2\,p^2}
+{\cal O}(a^3\,p^3)
\label{int6}
\eea
\bea
\hspace{-1cm}
\bullet\int_{-\pi}^\pi \frac{d^4k}{(2\pi)^4}\frac{\sum_\mu \hat{k}^4_\mu}
{16(\hat k^2)^2\, \widehat{k+a\,p}^2}
&=& 0.004050096698 - 0.000107954163\, a^2\,p^2 
+a^2\,\frac{\sum_{\mu} p^4_{\mu}}{1024\pi^2 \,p^2} \nonumber\\[1.5ex]
\hspace{-1cm}
&+&{\cal O}(a^4\,p^4)
\label{int7}\\[0.5ex]
\nonumber\\
\nonumber\\
\hspace{-1cm}
\bullet\int_{-\pi}^\pi \frac{d^4k}{(2\pi)^4}\frac{\kc_\mu\,\kc_\nu\,\kc_\rho\,\kc_\sigma}
{(\hat k^2)^2\, \widehat{k+a\,p}^2}&=&
0.001589337971\,(\delta_{\mu\nu }\,\delta_{\rho\sigma})_S -0.001675948042\,\delta_{\mu\nu\rho\sigma}\nonumber\\[1.5ex]
\hspace{-1cm}
&-&0.000372782983 (\delta_{\mu\nu\rho}\, a^2\,p_\mu \,p_\sigma)_S
-0.000062130497 (\delta_{\mu\nu}\, \delta_{\rho\sigma}\, a^2\,p_\mu^2)_S \nonumber\\[1.5ex]
\hspace{-1cm}
&+&\delta_{\mu\nu\rho\sigma}\left(\,0.000186391491\,a^2\,p^2 +0.000410290033\,a^2\,p_\mu^2\right)\nonumber\\[1.5ex]
\hspace{-1cm}
&+&(\delta_{\mu\nu} \,a^2\,p_\rho \,p_\sigma)_S\left(\,0.000227848225
-\frac{\ln(a^2 p^2)}{384\pi^2}\right)\nonumber\\[1.5ex]
\hspace{-1cm}
&+&(\delta_{\mu\nu }\, \delta_{\rho\sigma})_S \,a^2\,p^2\left(-0.000245852737
+\frac{\ln(a^2 p^2)}{768\pi^2}\right)\nonumber\\[1.5ex]
\hspace{-1cm}
&+&a^2\,\frac{p_\mu \,p_\nu \,p_\rho \,p_\sigma}{64\pi^2 \,p^2}+{\cal O}(a^4\,p^4)
\label{int8}\\[1.75ex]
\nonumber\\
\nonumber
\hspace{-1cm}
\bullet\int_{-\pi}^\pi \frac{d^4k}{(2\pi)^4}\frac{\kc_\nu\sum_\mu \hat{k}^4_\mu}
{16(\hat k^2)^2\, \widehat{k+a\,p}^2}&=&
a\,p_\nu \Bigl[-0.000800034900 +0.000069705553\,a^2\,p^2 \nonumber\\[0.85ex] 
\hspace{-1cm}
&&\phantom{a\,p_\nu \Bigl[}
+ 0.000107082394\,a^2\,p_\nu^2 -a^2\,\frac{\sum_\rho p^4_\rho}{1280\pi^2 \,p^2} \nonumber\\[0.85ex] 
\hspace{-1cm}
&&\phantom{a\,p_\nu \Bigl[}
-a^2\,\frac{\ln(a^2 p^2)}{2560\pi^2} \left(\frac{p^2}{2}-p_\nu^2\right) \Bigr]+{\cal O}(a^5\,p^5)
\label{int9}\\[0.5ex]
\nonumber\\
\nonumber\\
\hspace{-1cm}
\bullet\int_{-\pi}^\pi \frac{d^4k}{(2\pi)^4}\frac{\kc_\nu\,\kc_\rho
\sum_\mu \widehat{k_\mu{+}a\,p_\mu}^4}{16(\hat k^2)^2\, {({\widehat{k{+}a\,p}}^2)}^2}
&=&\delta_{\nu\rho}\Bigl[0.000400017450 -0.000034852777\,a^2\, p^2 
+a^2\,\frac{\sum_{\mu}p^4_{\mu}}{2560\pi^2 \,p^2}\nonumber\\[0.85ex]
\hspace{-1cm}
&&\phantom{\delta_{\nu\rho}\Bigl[}
+0.000105349447\,a^2\, p_\nu^2 +a^2\,\frac{\ln(a^2 p^2)}{5120\pi^2} (\frac{p^2}{2}-3p_\nu^2)\Bigr]\nonumber\\[0.85ex]
\hspace{-1cm}
&+&a^2\,p_\nu\,p_\rho\Bigl[\,0.000006643045 -\frac{p_\nu^2+p_\rho^2}{2560\pi^2 \,p^2}
+\frac{\sum_{\mu}p^4_{\mu}}{5120\pi^2 \,(p^2)^2}\nonumber\\[0.85ex]
\hspace{-1cm}
&&\phantom{a^2\,p_\nu\,p_\rho\Bigl[}
+\frac{\ln(a^2 p^2)}{5120\pi^2}\Bigr] +{\cal O}(a^4\,p^4)
\label{int10}
\eea

\bigskip\noindent
{\bf Acknowledgments:} This work is supported in part by the Research Promotion Foundation 
of Cyprus (Proposal Nr: ENI$\Sigma$X/0505/45, TEXN/0308/17).


\begin{table}
\begin{center}
\begin{minipage}{13cm}
\begin{tabular}{lr@{}lr@{}lr@{}l}
\hline
\hline
\multicolumn{1}{c}{Action}&
\multicolumn{2}{c}{$c_{0_{\Large{\phantom{A}}}}^{{\Large{\phantom{A}}}}$} &
\multicolumn{2}{c}{$c_1$} &
\multicolumn{2}{c}{$c_3$} \\
\hline
\hline
$\,\,$Plaquette               &  1&.0             &  0&                &  0&               \\
$\,\,$Symanzik                &  1&.6666667       & -0&.083333         &  0&               \\
$\,\,$TILW, $\beta c_0=8.60$  &  2&.3168064       & -0&.151791         & -0&.0128098$\,\,$  \\
$\,\,$TILW, $\beta c_0=8.45$  &  2&.3460240       & -0&.154846         & -0&.0134070$\,\,$  \\
$\,\,$TILW, $\beta c_0=8.30$  &  2&.3869776       & -0&.159128         & -0&.0142442$\,\,$  \\
$\,\,$TILW, $\beta c_0=8.20$  &  2&.4127840       & -0&.161827         & -0&.0147710$\,\,$  \\
$\,\,$TILW, $\beta c_0=8.10$  &  2&.4465400       & -0&.165353         & -0&.0154645$\,\,$  \\
$\,\,$TILW, $\beta c_0=8.00$  &  2&.4891712       & -0&.169805         & -0&.0163414$\,\,$  \\
$\,\,$Iwasaki                 &  3&.648           & -0&.331            &  0&                \\
$\,\,$DBW2                    & 12&.2688          & -1&.4086           &  0&                \\
\hline
\hline
\end{tabular}
\end{minipage}
\end{center}
\caption{Input parameters $c_0$, $c_1$, $c_3$.}\label{tab1}
%
\begin{center}
\begin{minipage}{13cm}
\begin{tabular}{lr@{}lr@{}lr@{}l}
\hline
\hline
\multicolumn{1}{c}{Action}&
\multicolumn{2}{c}{$\ve_{\phantom{A}}^{{(0,1)}^{\phantom{A}}}$}&
\multicolumn{2}{c}{$\ve^{(0,2)}$} &
\multicolumn{2}{c}{$\ve^{(0,3)}$} \\
\hline
\hline
$\,\,$Plaquette   &&16.6444139(2)    &-&2.24886853(7)   &-&1.39726711(7)          \\
$\,\,$Symanzik    &&13.02327272(7)   &-&2.01542504(4)   &-&1.24220271(2)          \\
$\,\,$TILW (8.45) &&10.82273528(9)   &-&1.84838009(3)   &-&1.13513794(1)          \\
$\,\,$TILW (8.00) &&10.45668970(6)   &-&1.81821854(5)   &-&1.11582732(3)          \\
$\,\,$Iwasaki     &&8.1165665(2)     &-&1.60101088(7)   &-&0.97320689(3)          \\
$\,\,$DBW2        &&2.9154231(2)     &-&0.96082198(5)   &-&0.56869876(4)$\,\,$    \\
\hline
\hline
\end{tabular}
\end{minipage}
\end{center}
\caption{The coefficients $\ve^{(0,i)}$ (Eq.~(\ref{propagator})) for
  different actions.}\label{tab2}
\end{table}
\begin{table}
\begin{center}
\begin{minipage}{13cm}
\begin{tabular}{lr@{}lr@{}lr@{}l}
\hline
\hline
\multicolumn{1}{c}{Action}&
\multicolumn{2}{c}{$\ve_{\phantom{A}}^{{(1,1)}^{\phantom{A}}}$}&
\multicolumn{2}{c}{$\ve^{(1,2)}$} &
\multicolumn{2}{c}{$\ve^{(1,3)}$} \\
\hline
\hline
$\,\,$Plaquette    &&12.8269254(2)    &-&5.20234231(6)    &-&0.08172763(4)           \\
$\,\,$Symanzik     &&10.69642966(8)   &-&4.7529781(1)     &-&0.075931174(1)          \\
$\,\,$TILW (8.45)  &&9.2865455(1)     &-&4.4186677(2)     &-&0.07160078(1)           \\
$\,\,$TILW (8.00)  &&9.0430829(2)     &-&4.35681290(3)    &-&0.070688697(3)          \\
$\,\,$Iwasaki      &&7.40724287(1)    &-&3.88883584(9)    &-&0.061025650(8)          \\
$\,\,$DBW2         &&3.0835163(2)     &-&2.2646221(1)     &-&0.03366740(1)$\,\,$     \\
\hline
\hline
\end{tabular}
\end{minipage}
\end{center}
\caption{The coefficients $\ve^{(1,i)}$ (Eq.~(\ref{propagator})) for
  different actions.}\label{tab3}
%
\begin{center}
\begin{minipage}{13cm}
\begin{tabular}{lr@{}lr@{}lr@{}l}
\hline
\hline
\multicolumn{1}{c}{Action}&
\multicolumn{2}{c}{$\ve_{\phantom{A}}^{{(2,1)}^{\phantom{A}}}$}&
\multicolumn{2}{c}{$\ve^{(2,2)}$} &
\multicolumn{2}{c}{$\ve^{(2,3)}$} \\
\hline
\hline
$\,\,$Plaquette    &-&4.74536466(2)   &&0.02028705(5)   &&0.10348577(3)   \\
$\,\,$Symanzik     &-&4.2478783(2)    &&0.05136635(6)   &&0.07865292(7)   \\
$\,\,$TILW (8.45)  &-&3.8139475(2)    &&0.05751390(9)   &&0.06651692(3)   \\
$\,\,$TILW (8.00)  &-&3.7342556(1)    &&0.05830392(9)   &&0.06444077(4)   \\
$\,\,$Iwasaki      &-&3.2018047(1)    &&0.08249970(7)   &&0.04192446(4)   \\
$\,\,$DBW2         &-&0.8678072(2)    &&0.1024452(2)    &-&0.00343999(2)$\,\,$\\
\hline
\hline
\end{tabular}
\end{minipage}
\end{center}
\caption{The coefficients $\ve^{(2,1)}\,-\,\ve^{(2,3)}$ (Eq.~(\ref{propagator})) for
  different actions.}\label{tab4}
%
\begin{center}
\begin{minipage}{13cm}
\begin{tabular}{lr@{}lr@{}lr@{}l}
\hline
\hline
\multicolumn{1}{c}{Action}&
\multicolumn{2}{c}{$\ve_{\phantom{A}}^{{(2,4)}^{\phantom{A}}}$}&
\multicolumn{2}{c}{$\ve^{(2,5)}$} &
\multicolumn{2}{c}{$\ve^{(2,6)}$} \\
\hline
\hline
$\,\,$Plaquette    &-&1.5048070(1)    &&0.70358496(5)   &&0.534320852(7)$\,\,$\\
$\,\,$Symanzik     &-&1.14716212(5)   &&0.65343092(3)   &&0.49783419(2)  \\
$\,\,$TILW (8.45)  &-&0.92583451(6)   &&0.62061757(5)   &&0.467966296(9) \\
$\,\,$TILW (8.00)  &-&0.8875297(1)    &&0.61441084(7)   &&0.462237852(9) \\
$\,\,$Iwasaki      &-&0.6202244(1)    &&0.55587473(6)   &&0.41846440(4)  \\
$\,\,$DBW2         &-&0.3202477(5)    &&0.34886590(2)   &&0.23968038(4)  \\
\hline
\hline
\end{tabular}
\end{minipage}
\end{center}
\caption{The coefficients $\ve^{(2,4)}\,-\,\ve^{(2,6)}$ (Eq.~(\ref{propagator})) for
  different actions.}\label{tab5}
\end{table}
\begin{table}
\begin{center}
\begin{minipage}{13cm}
\begin{tabular}{lr@{}lr@{}lr@{}l}
\hline
\hline
\multicolumn{1}{c}{Action}&
\multicolumn{2}{c}{$\tilde\ve^{{(0,1)}^{\phantom{A}}}_{1_{\phantom{A}}}$} &
\multicolumn{2}{c}{$\tilde\ve^{(1,1)}_1$} &
\multicolumn{2}{c}{$\tilde\ve^{(2,1)}_1$} \\
\hline
\hline
$\,\,$Plaquette     &9.&174787621(1)  &4.&5873938103(5)    &-1.&5291312701(2)     \\
$\,\,$Symanzik      &7.&071174701(5)  &3.&535587351(2)     &-1.&1785291169(8)     \\
$\,\,$TILW (8.45)   &5.&86097856(2)   &2.&930489282(8)     &-0.&976829761(3)      \\
$\,\,$TILW (8.00)   &5.&663791993(4)  &2.&831895997(2)     &-0.&9439653322(7)     \\
$\,\,$Iwasaki       &4.&423664730(5)  &2.&211832365(2)     &-0.&7372774550(8)$\,\,$\\
$\,\,$DBW2          &1.&86908767(4)   &0.&93454384(2)      &-0.&311514612(6)      \\
\hline
\hline
\end{tabular}
\end{minipage}
\end{center}
\caption{The coefficients $\tilde\ve_1^{(0,i)}$ (Eq.~(\ref{prop_d1})) for different actions.}\label{tab7}  
%
\begin{center}
\begin{minipage}{13cm}
\begin{tabular}{lr@{}lr@{}lr@{}l}
\hline
\hline
\multicolumn{1}{c}{Action}&
\multicolumn{2}{c}{$\tilde\ve^{{(0,1)}^{\phantom{A}}}_{2_{\phantom{A}}}$} &
\multicolumn{2}{c}{$\tilde\ve^{(1,1)}_2$} &
\multicolumn{2}{c}{$\tilde\ve^{(2,1)}_2$} \\
\hline
\hline
$\,\,$Plaquette    &7.&4696262(2)   &8.&2395316(2)    &-&3.21623339(2)    \\
$\,\,$Symanzik     &5.&95209802(7)  &7.&16084231(8)   &-&3.0693492(2)    \\
$\,\,$TILW (8.45)  &4.&96175672(9)  &6.&3560562(1)    &-&2.8371177(2)    \\
$\,\,$TILW (8.00)  &4.&79289770(6)  &6.&2111869(2)    &-&2.7902902(1)    \\
$\,\,$Iwasaki      &3.&6929018(2)   &5.&19541051(1)   &-&2.4645273(1)    \\
$\,\,$DBW2         &1.&0463355(2)   &2.&1489724(2)    &-&0.5562925(2)     \\
\hline
\hline
\end{tabular}
\end{minipage}
\end{center}
\caption{The coefficients $\tilde\ve_2^{(0,i)}$ (Eq.~(\ref{prop_d2})) for different actions.}\label{tab8}  
%
\begin{center}
\begin{minipage}{13cm}
\begin{tabular}{lr@{}lr@{}lr@{}l}
\hline
\hline
\multicolumn{1}{c}{Action}&
\multicolumn{2}{c}{$\ve_{S_{\phantom{A}}}^{{(0,1)}^{\phantom{A}}}$}&
\multicolumn{2}{c}{$\ve_S^{(0,2)}$} &
\multicolumn{2}{c}{$\ve_S^{(0,3)}$} \\
\hline
\hline
$\,\,$Plaquette    &&0.30799634(6)    &&9.9867847(2)    &&0.01688643(6)       \\
$\,\,$Symanzik     &&0.58345905(5)    &&8.8507071(1)    &-&0.12521126(5)      \\
$\,\,$TILW (8.45)  &&0.7049818(1)     &&8.0538938(2)    &-&0.20881716(3)      \\
$\,\,$TILW (8.00)  &&0.7195566(1)     &&7.9115477(2)    &-&0.22196498(3)      \\
$\,\,$Iwasaki      &&0.74092360(2)    &&6.9016820(2)    &-&0.29335071(4)      \\
$\,\,$DBW2         &-&0.0094234(5)    &&4.0385802(2)    &-&0.35869680(4)$\,\,$      \\
\hline
\hline
\end{tabular}
\end{minipage}
\end{center}
\caption{The coefficients $\ve_S^{(0,i)}$ (Eq.~(\ref{scalar})) for different actions.}\label{tab9}
\end{table}
\begin{table}
\begin{center}
\begin{minipage}{13cm}
\begin{tabular}{lr@{}lr@{}lr@{}l}
\hline
\hline
\multicolumn{1}{c}{Action}&
\multicolumn{2}{c}{$\ve_S^{(1,1)}$}&
\multicolumn{2}{c}{$\ve_S^{(1,2)}$} &
\multicolumn{2}{c}{$\ve_S^{(1,3)}$} \\
\hline
\hline
$\,\,$Plaquette    &&0.6586287(1)    &-&4.20298580(6)   &-&1.286053869(4)  \\
$\,\,$Symanzik     &&0.33939970(4)   &-&3.76353718(6)   &-&1.150059945(4)  \\
$\,\,$TILW (8.45)  &&0.1463203(2)    &-&3.42960982(2)   &-&1.054472092(1)  \\
$\,\,$TILW (8.00)  &&0.1155729(2)    &-&3.36704753(5)   &-&1.037165442(1)  \\
$\,\,$Iwasaki      &-&0.05097214(7)  &-&2.88571027(1)   &-&0.909503374(3)  \\
$\,\,$DBW2         &-&0.1248521(3)   &-&1.15247167(2)   &-&0.53943631(1)   \\
\hline
\hline
\end{tabular}
\end{minipage}
\end{center}
\caption{The coefficients $\ve_S^{(1,i)}$ (Eq.~(\ref{scalar})) for different actions.}\label{tab10}
%
\begin{center}
\begin{minipage}{13cm}
\begin{tabular}{lr@{}lr@{}lr@{}l}
\hline
\hline
\multicolumn{1}{c}{Action}&
\multicolumn{2}{c}{$\ve_{S_{\phantom{A}}}^{{(2,1)}^{\phantom{A}}}$}&
\multicolumn{2}{c}{$\ve_S^{(2,2)}$} &
\multicolumn{2}{c}{$\ve_S^{(2,3)}$} \\
\hline
\hline
$\,\,$Plaquette    &&2.60041308(7)    &-&4.15080331(7)    &&0.17641091(2)      \\
$\,\,$Symanzik     &&2.3547298(2)     &-&3.85277871(9)    &&0.196461884(5)     \\
$\,\,$TILW (8.45)  &&2.1881285(8)     &-&3.6249313(5)     &&0.21113016(1)      \\
$\,\,$TILW (8.00)  &&2.1605653(8)     &-&3.58171175(4)    &&0.21385016(2)      \\
$\,\,$Iwasaki      &&2.02123300(8)    &-&3.23459547(4)    &&0.234502732(7)$\,\,$\\
$\,\,$DBW2         &&2.3731619(3)     &-&1.9332087(1)     &&0.2953480(3)       \\
\hline
\hline
\end{tabular}
\end{minipage}
\end{center}
\caption{The coefficients $\ve_S^{(2,i)}$ (Eq.~(\ref{scalar})) for different actions.}\label{tab11}
%
\begin{center}
\begin{minipage}{16.5cm}
\begin{tabular}{lr@{}lr@{}lr@{}lr@{}l}
\hline
\hline
\multicolumn{1}{c}{Action}&
\multicolumn{2}{c}{$\ve_{P_{\phantom{A}}}^{{(0,1)}^{\phantom{A}}}$}&
\multicolumn{2}{c}{$\ve_P^{(0,2)}$} &
\multicolumn{2}{c}{$\ve_P^{(2,1)}$} &
\multicolumn{2}{c}{$\ve_P^{(2,2)}$}\\
\hline
\hline
$\,\,$Plaquette    &&9.95102761(8) &&3.43328275(3)   &&0.84419938(7)  &-&0.25823485(3)  \\
$\,\,$Symanzik     &&8.7100837(1)  &&2.98705498(3)   &&0.70640549(6)  &-&0.27556247(3)  \\
$\,\,$TILW (8.45)  &&7.84510495(6) &&2.67986902(3)   &&0.65030355(6)  &-&0.28812231(2)  \\
$\,\,$TILW (8.00)  &&7.6896423(1)  &&2.62578350(2)   &&0.64432843(6)  &-&0.29027771(3)  \\
$\,\,$Iwasaki      &&6.55611308(7) &&2.25383382(3)   &&0.66990790(5)  &-&0.30221183(3)  \\
$\,\,$DBW2         &&2.9781769(6)  &&1.24882665(4)   &&1.5569125(1)   &-&0.3362271(2)$\,\,$ \\
\hline
\hline
\end{tabular}
\end{minipage}
\end{center}
\caption{The coefficients $\ve_P^{(0,i)}$ and $\ve_P^{(2,i)}$
  (Eq.~(\ref{pseudoscalar})) for different actions.}\label{tab12}
\end{table}
\begin{table}
\begin{center}
\begin{minipage}{13cm}
\begin{tabular}{lr@{}lr@{}lr@{}l}
\hline
\hline
\multicolumn{1}{c}{Action}&
\multicolumn{2}{c}{$\ve_{V_{\phantom{A}}}^{{(0,1)}^{\phantom{A}}}$}&
\multicolumn{2}{c}{$\ve_V^{(0,2)}$} &
\multicolumn{2}{c}{$\ve_V^{(0,3)}$} \\
\hline
\hline
$\,\,$Plaquette    &&3.97338480(2)   &-&2.49669620(4)  &&0.85409908(1)   \\
$\,\,$Symanzik     &&3.57961385(3)   &-&2.21267683(2)  &&0.77806655(1)   \\
$\,\,$TILW (8.45)  &&3.32483844(4)   &-&2.01347343(2)  &&0.72217154(1)   \\
$\,\,$TILW (8.00)  &&3.28098129(5)   &-&1.97788691(3)  &&0.71193712(2)   \\
$\,\,$Iwasaki      &&2.98283189(2)   &-&1.72542048(4)  &&0.63679613(2)   \\
$\,\,$DBW2         &&2.25812410(4)   &-&1.00964505(3)  &&0.40188086(2)   \\
\hline
\hline
\end{tabular}
\end{minipage}
\end{center}
\caption{The coefficients $\ve_V^{(0,i)}$ (Eq.~(\ref{vector})) for different actions.}\label{tab13}
\end{table}
\begin{table}
\begin{center}
\begin{minipage}{13cm}
\begin{tabular}{lr@{}lr@{}lr@{}l}
\hline
\hline
\multicolumn{1}{c}{Action}&
\multicolumn{2}{c}{$\ve_{V_{\phantom{A}}}^{{(1,1)}^{\phantom{A}}}$}&
\multicolumn{2}{c}{$\ve_V^{(1,2)}$} &
\multicolumn{2}{c}{$\ve_V^{(1,3)}$} \\
\hline
\hline
$\,\,$Plaquette    &&2.7109817(1)    &-&1.84813992(2)  &-&0.39052850(2)    \\
$\,\,$Symanzik     &&2.09743725(3)   &-&1.51877201(8)  &-&0.385127257(2)   \\
$\,\,$TILW (8.45)  &&1.64290440(2)   &-&1.2579161(2)   &-&0.37793187(2)    \\
$\,\,$TILW (8.00)  &&1.55841933(4)   &-&1.20780827(3)  &-&0.3761606511(4)$\,\,$  \\
$\,\,$Iwasaki      &&0.9074321(1)    &-&0.80352187(4)  &-&0.356005234(3)   \\
$\,\,$DBW2         &-&1.4498098(4)   &&0.8826550(3)    &-&0.264655885(7)   \\
\hline
\hline
\end{tabular}
\end{minipage}
\end{center}
\caption{The coefficients $\ve_V^{(1,i)}$ (Eq.~(\ref{vector})) for different actions.}\label{tab14}
\end{table}
\begin{table}
\begin{center}
\begin{minipage}{13cm}
\begin{tabular}{lr@{}lr@{}lr@{}l}
\hline
\hline
\multicolumn{1}{c}{Action}&
\multicolumn{2}{c}{$\ve_{V_{\phantom{A}}}^{{(2,1)}^{\phantom{A}}}$} &
\multicolumn{2}{c}{$\ve_V^{(2,2)}$} &
\multicolumn{2}{c}{$\ve_V^{(2,3)}$}\\
\hline
\hline
$\,\,$Plaquette    &&1.5541024(2)      &&0.32907377(4)     &-&0.0060202576(6)$\,\,$\\
$\,\,$Symanzik     &&1.6762868(2)      &&0.22601986(5)     &&0.02822949(2)        \\
$\,\,$TILW (8.45)  &&1.63378530(3)     &&0.16772628(3)     &&0.04300929(5)        \\
$\,\,$TILW (8.00)  &&1.6190247(1)      &&0.15805313(3)     &&0.04550457(5)        \\
$\,\,$Iwasaki      &&1.4573118(1)      &&0.0858961(2)      &&0.07934994(2)        \\
$\,\,$DBW2         &-&1.1604825(4)     &-&0.0504803(3)     &&0.13992474(3)        \\
\hline
\hline
\end{tabular}
\end{minipage}
\end{center}
\caption{The coefficients $\ve_V^{(2,1)}\,-\,\ve_V^{(2,3)}$
  (Eq.~(\ref{vector})) for different actions.}\label{tab15}
\end{table}
\begin{table}
\begin{center}
\begin{minipage}{13cm}
\begin{tabular}{lr@{}lr@{}lr@{}l}
\hline
\hline
\multicolumn{1}{c}{Action}&
\multicolumn{2}{c}{$\ve_{V_{\phantom{A}}}^{{(2,4)}^{\phantom{A}}}$} &
\multicolumn{2}{c}{$\ve_V^{(2,5)}$} &
\multicolumn{2}{c}{$\ve_V^{(2,6)}$} \\
\hline
\hline
$\,\,$Plaquette    &&0.2500659(2)      &&0.8859920(1)	   &-&0.300364436(2)$\,\,$\\
$\,\,$Symanzik     &&0.0214112(1)      &&0.8342659(2)	   &-&0.28736163(1)     \\
$\,\,$TILW (8.45)  &-&0.1100958(1)     &&0.791026749(4)    &-&0.27444757(3)     \\ 
$\,\,$TILW (8.00)  &-&0.1318272(2)     &&0.78255494(3)     &-&0.27181092(1)     \\
$\,\,$Iwasaki      &-&0.2668492(1)     &&0.712786719(6)    &-&0.25078366(2)     \\
$\,\,$DBW2         &-&0.1528741(6)     &&0.42190739(4)     &-&0.13978037(7)     \\
\hline
\hline
\end{tabular}
\end{minipage}
\end{center}
\caption{The coefficients $\ve_V^{(2,4)}\,-\,\ve_V^{(2,6)}$
  (Eq.~(\ref{vector})) for different actions.}\label{tab17} 
\end{table}

\begin{table}
\begin{center}
\begin{minipage}{13cm}
\begin{tabular}{lr@{}lr@{}lr@{}l}
\hline
\hline
\multicolumn{1}{c}{Action}&
\multicolumn{2}{c}{$\ve_{V_{\phantom{A}}}^{{(2,7)}^{\phantom{A}}}$} &
\multicolumn{2}{c}{$\ve_V^{(2,8)}$} &
\multicolumn{2}{c}{$\ve_V^{(2,9)}$} \\
\hline
\hline
$\,\,$Plaquette    &&1.27887765(9)   &&0.27776135(2)  &-&0.35475044(2)    \\
$\,\,$Symanzik     &&1.03773908(9)   &&0.28969451(4)  &-&0.302816648(5)$\,\,$\\
$\,\,$TILW (8.45)  &&0.89400856(7)   &&0.2930984(2)   &-&0.25886703(3)    \\
$\,\,$TILW (8.00)  &&0.87034685(8)   &&0.29343883(9)  &-&0.25031691(5)    \\
$\,\,$Iwasaki      &&0.76263373(2)   &&0.29755270(5)  &-&0.184270928(8)   \\
$\,\,$DBW2         &&1.7371355(5)    &&0.2960594(1)   &&0.10831780(4)     \\
\hline
\hline
\end{tabular}
\end{minipage}
\end{center}
\caption{The coefficients $\ve_V^{(2,7)}\,-\,\ve_V^{(2,9)}$
  (Eq.~(\ref{vector})) for different actions.}\label{tab18} 
\end{table}
\begin{table}
\begin{center}
\begin{minipage}{16.5cm}
\begin{tabular}{lr@{}lr@{}lr@{}lr@{}l}
\hline
\hline
\multicolumn{1}{c}{Action}&
\multicolumn{2}{c}{$\ve_{A_{\phantom{A}}}^{{(0,1)}^{\phantom{A}}}$} &
\multicolumn{2}{c}{$\ve_A^{(1,1)}$} &
\multicolumn{2}{c}{$\ve_A^{(1,2)}$} &
\multicolumn{2}{c}{$\ve_A^{(1,3)}$} \\
\hline
\hline
$\,\,$Plaquette    &-&0.84813073(8)	  &&1.34274645(8)      &-&1.71809242(4)     &&0.130176166(7)	  \\
$\,\,$Symanzik     &-&0.48369852(8)	  &&0.92541220(1)      &-&1.54604828(4)     &&0.128375752(4)	  \\
$\,\,$TILW (8.45)  &-&0.2452231(1)	  &&0.64518173(2)      &-&1.42093097(4)     &&0.125977289(1)	  \\
$\,\,$TILW (8.00)  &-&0.20406156(8)     &&0.59652190(3)        &-&1.39831769(4)     &&0.125386884(3)      \\
$\,\,$Iwasaki      &&0.0752372(1)       &&0.2684958(1)         &-&1.238019617(7)    &&0.1186684108(9)$\,\,$\\
$\,\,$DBW2         &&0.7643240(1)       &-&0.56650487(5)       &-&0.75581589(7)     &&0.088218628(3)      \\
\hline
\hline
\end{tabular}
\end{minipage}
\end{center}
\caption{The coefficients $\ve_A^{(0,1)}$ and $\ve_A^{(1,i)}$
  (Eq.~(\ref{axial})) for different actions.}\label{tab19} 
\end{table}

\begin{table}
\begin{center}
\begin{minipage}{13cm}
\begin{tabular}{lr@{}lr@{}lr@{}l}
\hline
\hline
\multicolumn{1}{c}{Action}&
\multicolumn{2}{c}{$\ve_{A_{\phantom{A}}}^{{(2,1)}^{\phantom{A}}}$} &
\multicolumn{2}{c}{$\ve_A^{(2,2)}$} &
\multicolumn{2}{c}{$\ve_A^{(2,3)}$} \\
\hline
\hline
$\,\,$Plaquette    &&0.3879068(1)      &&1.85116980(8)   &-&0.093094486(8)$\,\,$    \\
$\,\,$Symanzik     &&0.29616583(7)     &&1.7629637(2)    &-&0.11136345(3)     \\
$\,\,$TILW (8.45)  &&0.2483248(2)      &&1.65782471(1)   &-&0.118658311(8)    \\
$\,\,$TILW (8.00)  &&0.2378781(2)      &&1.63840648(4)   &-&0.11982438(1)     \\
$\,\,$Iwasaki      &&0.05917686(4)     &&1.5707047(2)    &-&0.13932655(1)     \\
$\,\,$DBW2         &-&2.2341918(4)     &&1.22932319(6)   &-&0.17119304(8)     \\
\hline
\hline
\end{tabular}
\end{minipage}
\end{center}
\caption{The coefficients $\ve_A^{(2,1)}\,-\,\ve_A^{(2,3)}$
  (Eq.~(\ref{axial})) for different actions.}\label{tab20} 
\end{table}
\begin{table}
\begin{center}
\begin{minipage}{13cm}
\begin{tabular}{lr@{}lr@{}lr@{}l}
\hline
\hline
\multicolumn{1}{c}{Action}&
\multicolumn{2}{c}{$\ve_{A_{\phantom{A}}}^{{(2,4)}^{\phantom{A}}}$} &
\multicolumn{2}{c}{$\ve_A^{(2,5)}$} &
\multicolumn{2}{c}{$\ve_A^{(2,6)}$} \\
\hline
\hline
$\,\,$Plaquette    &&1.6350438(1)         &-&1.59945524(6)    &&0.333900263(8)$\,\,$\\
$\,\,$Symanzik     &&1.3008790(1)         &-&1.48761993(3)    &&0.314172576(5) \\
$\,\,$TILW (8.45)  &&1.0461303(2)         &-&1.39361896(4)    &&0.297787700(5) \\
$\,\,$TILW (8.00)  &&0.9998744(1)         &-&1.37577372(4)    &&0.29455820(2)  \\
$\,\,$Iwasaki      &&0.6845753(1)         &-&1.24800562(3)    &&0.26827353(2)  \\
$\,\,$DBW2         &&0.0967251(2)         &-&0.735419342(8)   &&0.14738921(4)  \\
\hline
\hline
\end{tabular}
\end{minipage}
\end{center}
\caption{The coefficients $\ve_A^{(2,4)}\,-\,\ve_A^{(2,6)}$
  (Eq.~(\ref{axial})) for different actions.}\label{tab21} 
\end{table}
\begin{table}
\begin{center}
\begin{minipage}{13cm}
\begin{tabular}{lr@{}lr@{}lr@{}l}
\hline
\hline
\multicolumn{1}{c}{Action}&
\multicolumn{2}{c}{$\ve_{A_{\phantom{A}}}^{{(2,7)}^{\phantom{A}}}$} &
\multicolumn{2}{c}{$\ve_A^{(2,8)}$} &
\multicolumn{2}{c}{$\ve_A^{(2,9)}$}  \\
\hline
\hline
$\,\,$Plaquette    &&0.41758917(4)     &&0.395847810(9)  &&0.31972188(2) \\
$\,\,$Symanzik     &&0.596637529(2)    &&0.33473715(4)   &&0.27870681(2) \\
$\,\,$TILW (8.45)  &&0.73021636(8)     &&0.29171961(4)   &&0.24115534(2) \\
$\,\,$TILW (8.00)  &&0.75716237(6)     &&0.28297665(3)   &&0.233647603(8 \\
$\,\,$Iwasaki      &&1.05772129(4)     &&0.18672220(2)   &&0.17428813(3) \\
$\,\,$DBW2         &&3.4449465(4)      &-&0.22085461(4)  &-&0.10748502(5)$\,\,$ \\
\hline
\hline
\end{tabular}
\end{minipage}
\end{center}
\caption{The coefficients $\ve_A^{(2,7)}\,-\,\ve_A^{(2,9)}$
  (Eq.~(\ref{axial})) for different actions.}\label{tab22} 
\end{table}
\begin{table}
\begin{center}
\begin{minipage}{13cm}
\begin{tabular}{lr@{}lr@{}lr@{}l}
\hline
\hline
\multicolumn{1}{c}{Action}&
\multicolumn{2}{c}{$\ve_{T_{\phantom{A}}}^{{(0,1)}^{\phantom{A}}}$} &
\multicolumn{2}{c}{$\ve_T^{(0,2)}$} &
\multicolumn{2}{c}{$\ve_T^{(0,3)}$} \\
\hline
\hline
$\,\,$Plaquette     &&0.37366536(7)	  &-&1.66446414(3)	&-&0.5750281973(1)$\,\,$\\
$\,\,$Symanzik      &&0.51501972(4)  	  &-&1.47511786(3)	&-&0.4769739579(4)      \\
$\,\,$TILW (8.45)   &&0.62806240(5)	  &-&1.34231565(2)	&-&0.411841977(2)	\\
$\,\,$TILW (8.00)   &&0.64974666(4)       &-&1.31859128(1)  	&-&0.4006364262(1)      \\
$\,\,$Iwasaki       &&0.82253993(3)       &-&1.15028034(3)  	&-&0.3267471901(5)      \\
$\,\,$DBW2          &&1.5201736(4)        &-&0.67309671(3)  	&-&0.1483549734(1)      \\
\hline
\hline
\end{tabular}
\end{minipage}
\end{center}
\caption{The coefficients $\ve_T^{(0,i)}$ (Eq.~(\ref{tensor})) for different actions.}\label{tab23}
\end{table}
\begin{table}
\begin{center}
\begin{minipage}{13cm}
\begin{tabular}{lr@{}lr@{}lr@{}l}
\hline
\hline
\multicolumn{1}{c}{Action}&
\multicolumn{2}{c}{$\ve_{T_{\phantom{A}}}^{{(1,1)}^{\phantom{A}}}$} &
\multicolumn{2}{c}{$\ve_T^{(1,2)}$} &
\multicolumn{2}{c}{$\ve_T^{(1,3)}$} \\
\hline
\hline
$\,\,$Plaquette     &-&4.05372833(7)	  &&1.866287582(5)	&-&0.8573692476(6)$\,\,$ \\
$\,\,$Symanzik      &-&3.0228493(1)	  &&1.59558642(1)	&-&0.7667066321(8)       \\
$\,\,$TILW (8.45)   &-&2.28808611(8)      &&1.39406610(3)       &-&0.7029813946(7)       \\
$\,\,$TILW (8.00)   &-&2.15494125(2)      &&1.357142525(7)      &-&0.691443627(2)        \\
$\,\,$Iwasaki       &-&1.17592792(3)      &&1.087913642(4)      &-&0.6063355831(7)       \\
$\,\,$DBW2          &&2.0163147(3)        &&0.15488056(9)       &-&0.359624204(3)        \\
\hline
\hline
\end{tabular}
\end{minipage}
\end{center}
\caption{The coefficients $\ve_T^{(1,i)}$ (Eq.~(\ref{tensor})) for different actions.}\label{tab24}
\end{table}
\begin{table}
\begin{center}
\begin{minipage}{13cm}
\begin{tabular}{lr@{}lr@{}lr@{}l}
\hline
\hline
\multicolumn{1}{c}{Action}&
\multicolumn{2}{c}{$\ve_{T_{\phantom{A}}}^{{(2,1)}^{\phantom{A}}}$} &
\multicolumn{2}{c}{$\ve_T^{(2,2)}$} &
\multicolumn{2}{c}{$\ve_T^{(2,3)}$} \\
\hline
\hline
$\,\,$Plaquette     &&2.3328621(2)    &-&1.52209604(8)       &&0.23683195(1)$\,\,$ \\
$\,\,$Symanzik      &&2.4912319(2)    &-&1.53694399(3)       &&0.26295051(2)     \\
$\,\,$TILW (8.45)   &&2.4578347(1)    &-&1.49009843(6)       &&0.26767225(1)     \\
$\,\,$TILW (8.00)   &&2.44550431(4)   &-&1.48035333(8)       &&0.26834751(3)     \\
$\,\,$Iwasaki       &&2.3441345(2)    &-&1.4848088(1)        &&0.30172406(2)     \\
$\,\,$DBW2          &&1.3013094(2)    &-&1.2798033(3)        &&0.3475909(1)      \\
\hline
\hline
\end{tabular}
\end{minipage}
\end{center}
\caption{The coefficients $\ve_T^{(2,1)}\,-\,\ve_T^{(2,3)}$
  (Eq.~(\ref{tensor})) for different actions.}\label{tab25} 
\end{table}
\begin{table}
\begin{center}
\begin{minipage}{13cm}
\begin{tabular}{lr@{}lr@{}lr@{}l}
\hline
\hline
\multicolumn{1}{c}{Action}&
\multicolumn{2}{c}{$\ve_{T_{\phantom{A}}}^{{(2,4)}^{\phantom{A}}}$} &
\multicolumn{2}{c}{$\ve_T^{(2,5)}$} &
\multicolumn{2}{c}{$\ve_T^{(2,6)}$} \\
\hline
\hline
$\,\,$Plaquette   &-&2.02795509(9) 	 &&0.11808647(1)   &&0.07250824(3)       \\
$\,\,$Symanzik    &-&1.55221265(9) 	 &&0.04504264(4)   &&0.07020813(1)       \\
$\,\,$TILW (8.45) &-&1.2361662(1)        &-&0.00137858(3)  &&0.06309676(1)       \\
$\,\,$TILW (8.00) &-&1.17754207(2)       &-&0.0104623(1)   &&0.06169762(2)       \\
$\,\,$Iwasaki     &-&0.6509124(1)        &-&0.11083048(4)  &&0.06071268(8)       \\
$\,\,$DBW2        &&1.4802111(2)         &-&0.51691409(6)  &&0.03446776(7)$\,\,$ \\
\hline
\hline
\end{tabular}
\end{minipage}
\end{center}
\caption{The coefficients $\ve_T^{(2,4)}\,-\,\ve_T^{(2,6)}$
  (Eq.~(\ref{tensor})) for different actions.}\label{tab26} 
\end{table}
\begin{table}
\begin{center}
\begin{minipage}{13cm}
\begin{tabular}{lr@{}lr@{}lr@{}lr@{}l}
\hline
\hline
\multicolumn{1}{c}{Action}&
\multicolumn{2}{c}{$\ve_{T_{\phantom{A}}}^{{(2,7)}^{\phantom{A}}}$} &
\multicolumn{2}{c}{$\ve_T^{(2,8)}$} &
\multicolumn{2}{c}{$\ve_T^{(2,9)}$} \\
\hline
\hline
$\,\,$Plaquette   	  &&0.3932905(2)    &&1.10184617(6)   &-&0.02744360(1)      \\
$\,\,$Symanzik    	  &&0.1467998(1)    &&1.03762644(2)   &-&0.03500219(3)      \\
$\,\,$TILW (8.45)         &-&0.0094312(1)   &&0.97633518(7)   &-&0.038311849(6)$\,\,$\\
$\,\,$TILW (8.00)         &-&0.0364177(1)   &&0.96442476(2)   &-&0.03892456(3)      \\
$\,\,$Iwasaki             &-&0.2049940(1)   &&0.88259379(5)   &-&0.04896801(2)      \\
$\,\,$DBW2                &-&0.3118850(6)   &&0.51292384(2)   &-&0.07146761(1)      \\
\hline
\hline
\end{tabular}
\end{minipage}
\end{center}
\caption{The coefficients $\ve_T^{(2,7)}\,-\,\ve_T^{(2,9)}$
  (Eq.~(\ref{tensor})) for different actions.}\label{tab27} 

\vspace{0.35cm}
\begin{center}
\begin{minipage}{13cm}
\begin{tabular}{lr@{}lr@{}lr@{}l}
\hline
\hline
\multicolumn{1}{c}{Action}&
\multicolumn{2}{c}{$\ve_{{T'}_{\phantom{A}}}^{{(2,1)}^{\phantom{A}}}$} &
\multicolumn{2}{c}{$\ve_{T'}^{(2,4)}$} &
\multicolumn{2}{c}{$\ve_{T'}^{(2,7)}$} \\
\hline
\hline
$\,\,$Plaquette     &&0.00047095(5)     &-&0.30537822(7)   &&1.4070324(1)        \\
$\,\,$Symanzik      &-&0.26900984(7)    &-&0.67000961(5)   &&1.0574111(1)        \\
$\,\,$TILW (8.45)   &-&0.31308657(6)    &-&0.90858179(3)   &&0.7651952(1)        \\
$\,\,$TILW (8.00)   &-&0.31678921(7)    &-&0.95117312(1)   &&0.7107479(2)        \\ 
$\,\,$Iwasaki       &-&0.45213470(9)    &-&1.24108756(9)   &&0.3465297(2)        \\
$\,\,$DBW2          &-&0.8461093(2)     &-&1.9354110(1)    &-&0.6289363(3)$\,\,$ \\
\hline
\hline
\end{tabular}
\end{minipage}
\end{center}
\caption{The coefficients $\ve_{T'}^{(2,1)},\,\ve_{T'}^{(2,4)},\,\ve_{T'}^{(2,7)}$
  (Eq.~(\ref{tensorprime})) for different actions.}\label{tab30} 
\vspace{0.35cm}
\begin{center}
\begin{minipage}{10cm}
\begin{tabular}{lr@{}lr@{}l}
\hline
\hline
\multicolumn{1}{c}{Action}&
\multicolumn{2}{c}{$\ve_{{T'}_{\phantom{A}}}^{{(2,8)}^{\phantom{A}}}$} &
\multicolumn{2}{c}{$\ve_{T'}^{(2,9)}$}  \\
\hline
\hline
$\,\,$Plaquette     &&0.28175492(1) 	&&0.054718244(7)    \\
$\,\,$Symanzik      &&0.24663315(4) 	&&0.06136902(2)     \\
$\,\,$TILW (8.45)   &&0.2319752(1)      &&0.06397589(2)     \\
$\,\,$TILW (8.00)   &&0.22947913(7)     &&0.064400395(7)$\,\,$\\
$\,\,$Iwasaki       &&0.19560470(5)     &&0.07153771(1)     \\
$\,\,$DBW2          &&0.13147908(9)     &&0.08509400(2)     \\
\hline
\hline
\end{tabular}
\end{minipage}
\end{center}
\caption{The coefficients $\ve_{T'}^{(2,8)},\,\ve_{T'}^{(2,9)}$
  (Eq.~(\ref{tensorprime})) for different actions.}\label{tab31} 
\end{table}

\end{document}